\documentclass[fleqn,usenatbib]{mnras}

\usepackage{newtxtext,newtxmath}

\usepackage[T1]{fontenc}

\DeclareRobustCommand{\VAN}[3]{#2}
\let\VANthebibliography\thebibliography
\def\thebibliography{\DeclareRobustCommand{\VAN}[3]{##3}\VANthebibliography}

\usepackage{epsfig}
\usepackage{amsmath}
\usepackage{color}
\usepackage{xcolor}
\usepackage{booktabs}
\usepackage{pdflscape}
\usepackage{textcomp}

\usepackage{soul}
\usepackage{aas_macros}  
\usepackage{chngcntr}
\usepackage{array}
\newcolumntype{P}[1]{>{\raggedright\arraybackslash}p{#1}}
\graphicspath{{./}}

  \newcommand{\Teff}{\mbox{\,\em T$_{\rm eff}$}}         

  \newcommand{\kmsec}{\,\mbox{$\mbox{km}\,\mbox{s}^{-1}$}}    
  \def\simge{\mathrel{\raise1.16pt\hbox{$>$}\kern-7.0pt
    \lower3.06pt\hbox{{$\scriptstyle \sim$}}}}           
  \def\simle{\mathrel{\raise1.16pt\hbox{$<$}\kern-7.0pt
    \lower3.06pt\hbox{{$\scriptstyle \sim$}}}}           
\title[SALT hot subdwarfs: classification]{The SALT survey of helium-rich hot subdwarfs: final sample and classification}
\author[Jeffery et al.]{C. S. Jeffery$^{1}$, M. Dorsch$^{2}$, A. Philip Monai$^{1,3}$, E. J. Snowdon$^{1,3,4}$, I. Monageng$^{5,6}$, and B. Miszalski$^{7}$ \\
$^1$Armagh Observatory and Planetarium, College Hill, Armagh BT61 9DG, United Kingdom\\
$^2$Institut für Physik und Astronomie, Universität Potsdam, Haus 28, Karl-Liebknecht-Str.\ 24/25, 14476 Potsdam, Germany\\
$^3$School of Mathematics and Physics, Queen’s University Belfast, Belfast BT7 1NN, UK\\
$^4$Astrophysics Research Cluster, School of Mathematical and Physical Sciences, University of Sheffield, Sheffield S3 7RH, UK\\
$^5$Department of Astronomy, University of Cape Town, Private Bag X3, Rondebosch 7701, South Africa\\
$^6$South African Astronomical Observatory, PO Box 9, Observatory Rd., Observatory 7935, Cape Town, South Africa\\
$^7$Australian Astronomical Optics, Faculty of Science and Engineering, Macquarie University, North Ryde, NSW 2113, Australia
}

\begin{document}

\date{Accepted \ldots. Received \ldots; in original form \ldots}

\pagerange{\pageref{firstpage}--\pageref{lastpage}} \pubyear{2020}

\maketitle

\label{firstpage}

\begin{abstract}
A medium-resolution spectroscopic survey of helium-rich hot subdwarfs has been carried out using the Southern African Large Telescope (SALT).
Objectives include the discovery of exotic hot subdwarfs, resolving distinct subclasses, identifying evolutionary sequences, and establishing the past and future histories of many of these  unusual stars. 
This paper extends the sample described by \citet{jeffery21a} from 100 to {697} stars. 
It describes the selection criteria and presents spectral classifications based on the MK-like Drilling system. 
The sample includes 283 extremely helium-rich hot subdwarfs, {17} extreme helium stars, {110} intermediate helium-rich hot subdwarfs, as well as 21 helium-rich stars of other types. 
It now represents the largest homogeneous sample of both "normal" He-sdOs and "luminous" or "hot" He-sdOs.  
Interesting stars discovered include magnetic hot subdwarfs, extremely hot pre-white dwarfs and hot subdwarfs, including hot subdwarfs showing \ion{N}{v} emission, one short-period binary, new extreme helium stars and several double-subdwarf candidates. 
The data form the basis for kinematic and model atmosphere analyses to follow.  
\end{abstract}


\section{Introduction}

Hot subluminous stars are faint blue stars which lie beneath the main sequence on a classical Hertzsprung-Russell or colour-magnitude diagram \citep{humason47}. 
They include low-mass stars with a wide range of radii and surface chemistries and represent a late stage in a variety of different stellar evolution pathways. 
Hydrogen-rich subdwarf B (sdB) stars are often characterized as the cores of red giants stripped by a binary companion, or otherwise as core helium-burning extreme horizontal-branch stars \citep{han02}. 
The rarer sdOB and sdO stars, which lie on or around the helium main sequence, may have evolved from sdB stars or be low-mass stars contracting to become helium white dwarfs, while more luminous sdO stars may be post-AGB stars contracting to become CO white dwarfs. 
The majority of hot subdwarfs ($\approx 90 \%$)  have hydrogen-rich surfaces, with helium-to-hydrogen number fraction $y \equiv n_{\rm He}/n_{\rm H} <0.1$.  
A large part of the minority ($\approx 9 \%$ total) with helium-rich surfaces ($y > 10$) is more difficult to explain with binary mass-transfer models, but could provide evidence for a population of merged double white dwarfs \citep{iben90,saio00,zhang12a,schwab18}; these are the He-sdO and He-sdB stars. 
The remaining $\approx 1 \%$ feature intermediate  helium surfaces ($0.1 < y < 10$) and are extremely hard to explain, particularly as they include stars with exotic surface chemistries \citep{naslim11,jeffery19b}, strong magnetic fields \citep{dorsch24}, and exotic binaries \citep{snowdon23b}; these are the iHe-sdOB stars. 

Up to 2015, an impediment to unravelling the stellar evolution pathways that connect these various groups of stars was the relatively small number of hydrogen-deficient hot subdwarfs which had been studied in detail. 
57 hot subdwarfs were identified as helium-rich in the low-dispersion Edinburgh Cape (EC) survey of faint blue stars \citep[{\it e.g. \rm}][]{Cat.EC1}; some had classifications similar to those of extreme helium stars \citep[][D13]{drilling13}. 
A strong early motivation was to demonstrate whether some helium-rich subdwarfs could be connected to the extreme helium stars, either by a single or more than one pathway.  
Thus both hydrogen-deficient subdwarfs from the EC survey and extreme helium stars were included in an intermediate-dispersion survey carried out with the Southern African Large Telescope \citep[SALT,][]{jeffery21a}, together with 43 other potential H-deficient stars identified from other surveys.
This SALT survey comprised 87 He-sdOs, a substantial number at the time, as well as 5 extreme helium stars -- not quite enough to assess population statistics. 
The discovery of six new iHe-sdOBs only raised further questions; for instance, the case of the lead-rich binary EC\,22536–5304 \citep[][]{jeffery19b,dorsch21}, which was recently proposed to have achieved its heavy metal enrichment through self-synthesis \citep{battich25}. 
At the same time, the number of newly identified He-sdBs \citep[{\it e.g.\rm }][]{jeffery17b,jeffery24} remained intriguing but limited. 

Meanwhile, the {\it Gaia} spacecraft provided a new way to identify hot subdwarfs by providing both absolute magnitude and colour. 
This opened a way to significantly increase the sample of hydrogen-deficient subdwarfs. The aim of this work is to present an overview of a new SALT survey, based on targets selected using \textit{Gaia} parallaxes. 
Section \ref{s:sample} describes how the sample was selected, and how the observations were carried out. 
Section \ref{s:classes}  provides spectral classifications and  section \ref{s:stars} identifies stars of particular interest. 
More detailed analyses of the SALT spectra presented here will be carried out in separate studies. 
Philip Monai et al.\ (2025; submitted) will discuss an automated clustering and kinematical analysis of the sample.   
Dorsch et al.\ (2025; in prep) will present spectral analyses and atmosphere properties, and will discuss evolutionary connections between various classes within the sample.   

\begin{figure}
\begin{center}
\includegraphics[width=1.0\linewidth]{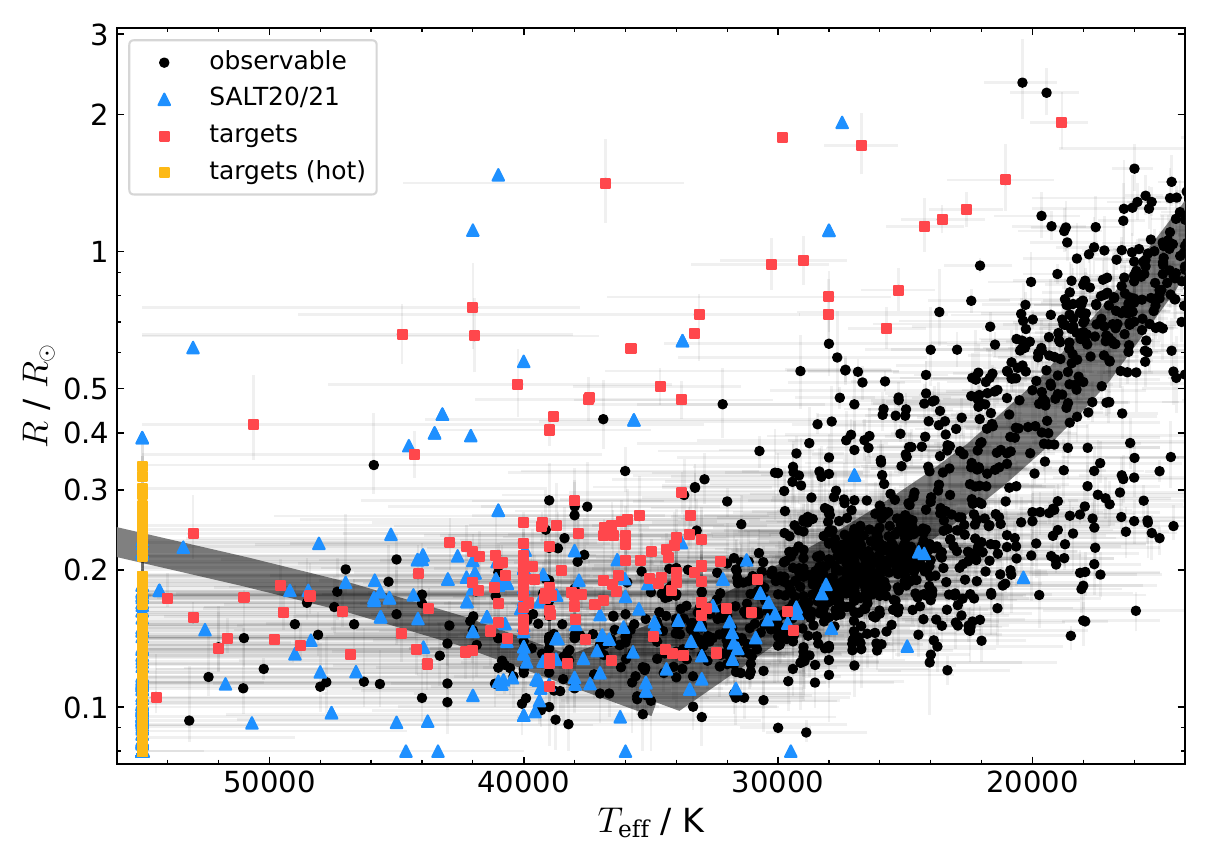}
\vspace{-10pt}
\caption{Sample selection using provisional surface temperatures and radii for 9979 hot subdwarfs based on spectral energy distributions and \textit{Gaia} distances. 
Stars observable with SALT are shown in black. 
Blue triangles show hydrogen-deficient subdwarfs identified observed in SALT1. 
Red and orange squares identify the SALT target list used for this campaign. 
Error bars indicate the considerable uncertainties that arise from SED fitting. 
The SED method is unable to constrain effective temperatures above 50\,000\,K.
Measurements above this value (in orange) should be simply interpreted as being $>50\,000$\,K or ``very hot'' rather than being assigned a numerical value. 
The gray bands mark the helium main sequence of \citet{paczynski71} and the \citet{dorman93} horizontal branch.} 
\label{f:sed_R}
\end{center}
\end{figure}

\section{Observations}
\label{s:sample}

\subsection{Target selection}

The original scope of the survey selected targets by their designation as a helium-rich hot subdwarf or extreme helium star, or their classification as He-sdB, He-sdOB, He-sdO (or sdOD) in one or more spectroscopic catalogues or equivalent. 
\citet{jeffery21a} reported results based on the first 100 stars observed during 2018 and 2019.
By the end of 2022, about 200 additional stars had been observed in the same way \citep{jeffery23c}.  
In the third phase of the survey, target selection was based on an extensive \textit{Gaia}-selected sample of hot subdwarfs \citep{culpan22} from which provisional effective temperatures, radii and luminosities could be derived directly from spectral energy distributions (SEDs) and \textit{Gaia} DR3 distances \citep{gaia23.dr3}. 
{For a detailed description of the SED method, see \citet{heber18}. }
Figure \ref{f:sed_R} shows these distributions. 
Stars with spectra from SALT or ESO instruments, or with spectroscopic classifications in the hot subdwarf catalogue \citep{culpan22} were removed. 
The \citet{jeffery21a} sample (in blue) lies principally at large radius or at high temperature relative to the dominant population of hot subdwarfs on the extreme horizontal branch {(EHB)} or classical hot subdwarf sequence (in black).
Additional targets for the present sample (in red) were selected as stars observable with SALT whose radii are consistent with being larger than 0.4\,$R_\odot$ and/or whose temperatures satisfy $T_{\rm eff} > 32000$\,K. 
{These criteria select against classical sdB stars with hydrogen-rich surfaces which lie on the EHB. Selected stars are therefore more likely to be helium-rich. }
A faint limit $m_{\rm G} \lesssim 16$ corresponds to a SALT observation of $\approx 1$h,  
and resulted in a further 388 stars being added to the sample. 
Note that the SED method does not constrain the surface composition of the stars; as a result, the new targets include both helium-rich and helium-poor stars. 

\subsection{SALT/RSS}
Observations were obtained with the SALT Robert Stobie Spectrograph \citep[RSS: resolution $R\approx 3\,600$,][]{burgh03,kobulnicky03} between 2017 and 2025 February.
Double exposures were taken at two different grating angles to provide continuous spectra in the wavelength range 3850 -- 5150 \AA. 
{Data obtained from observations up to 2020 were reduced using the {\sc pysalt} package \citep{crawford10} as described by \citet{jeffery21a}. 
Due to deprecation of some {\sc pysalt} dependencies, subsequent data reduction was carried out using a modification of the Titus Saures Rex pipeline\footnote{https://github.com/NaomiTitus/Titus\_Saures\_Rex\_Spectroscopic\_Routines}, as described in \citep{snowdon23.phd}. 
The replacement pipeline was selected due to its similarity to the {\sc pysalt} workflow, ensuring compatibility between pre- and post-2020 reductions.}

The one-dimensional wavelength-calibrated and sky-subtracted spectra were rectified using low-order polynomials fitted to regions of continuum. 
The three segments from both observations at both grating angles were merged  using weights based on the number of photons detected in each segment. 
Residual cosmic-ray spikes were removed by co-incidence matching between pairs of spectra wherever possible. 
The wavelengths of each spectrum were adjusted to correct for Earth motion around the solar system barycentre. 

125 stars were observed on more than one occasion, either to accumulate sufficient signal-to-noise in the case of faint stars or poor conditions, or specifically to check for radial-velocity variability. 

\begin{figure*}
\begin{center}
\includegraphics[clip, width=0.98\linewidth]{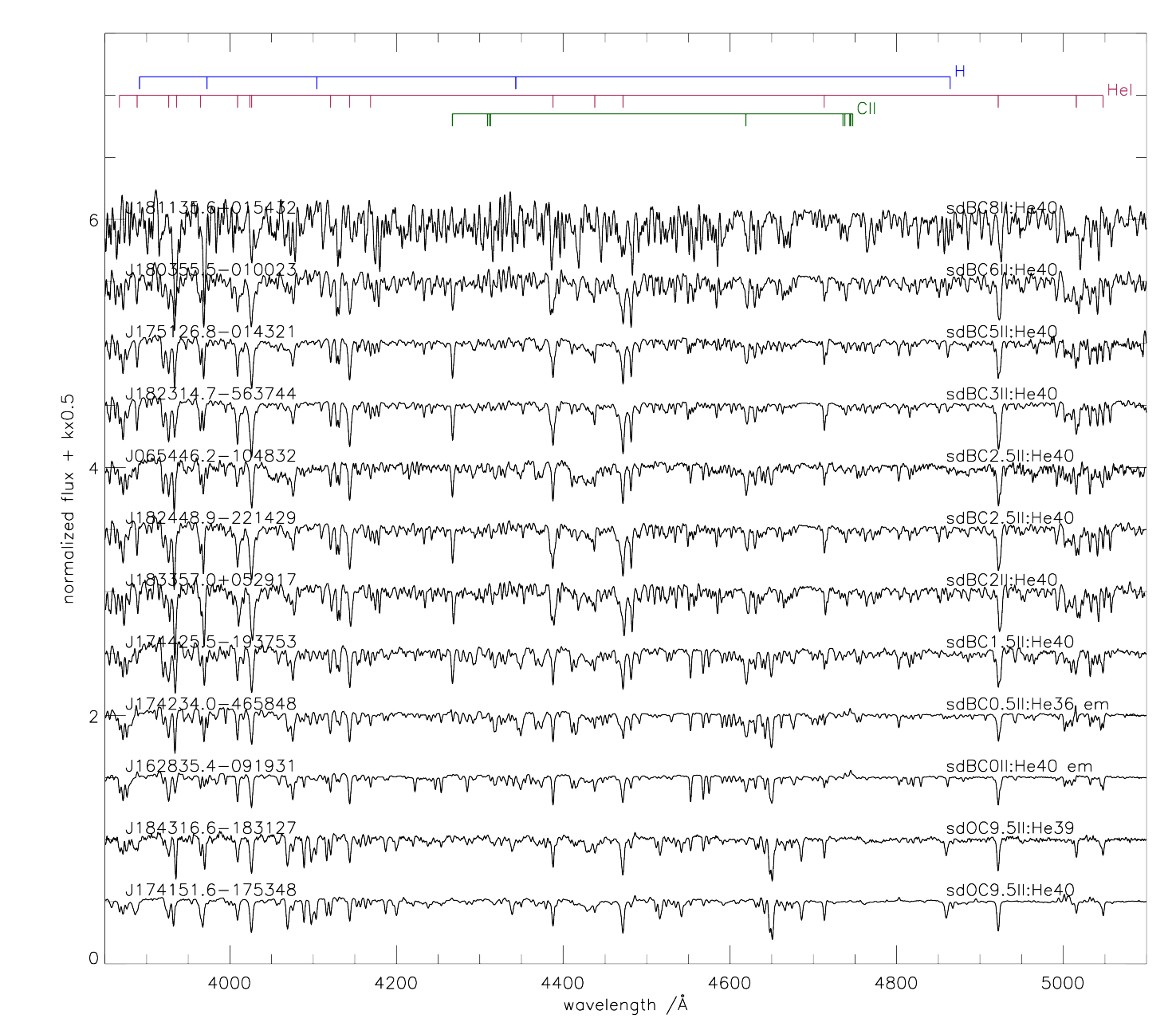}
\caption{SALT/RSS spectra of extreme helium stars from the narrow \ion{He}{i} line sequence. Spectra are labelled with their SALT identifiers (Table \ref{a:classes}ff.). Counting from the top, selected familiar EHes are:  2: NO\,Ser, 4: PV\,Tel = HD 168476 \citep{thackeray54}, 10: V2205\,Oph, 12: V2076\,Oph = HD\,160641 \citep{bidelman52}.} 
\label{f:ehe_lo}
\end{center}
\end{figure*}

\begin{figure*}
\begin{center}
\includegraphics[clip, width=0.98\linewidth]{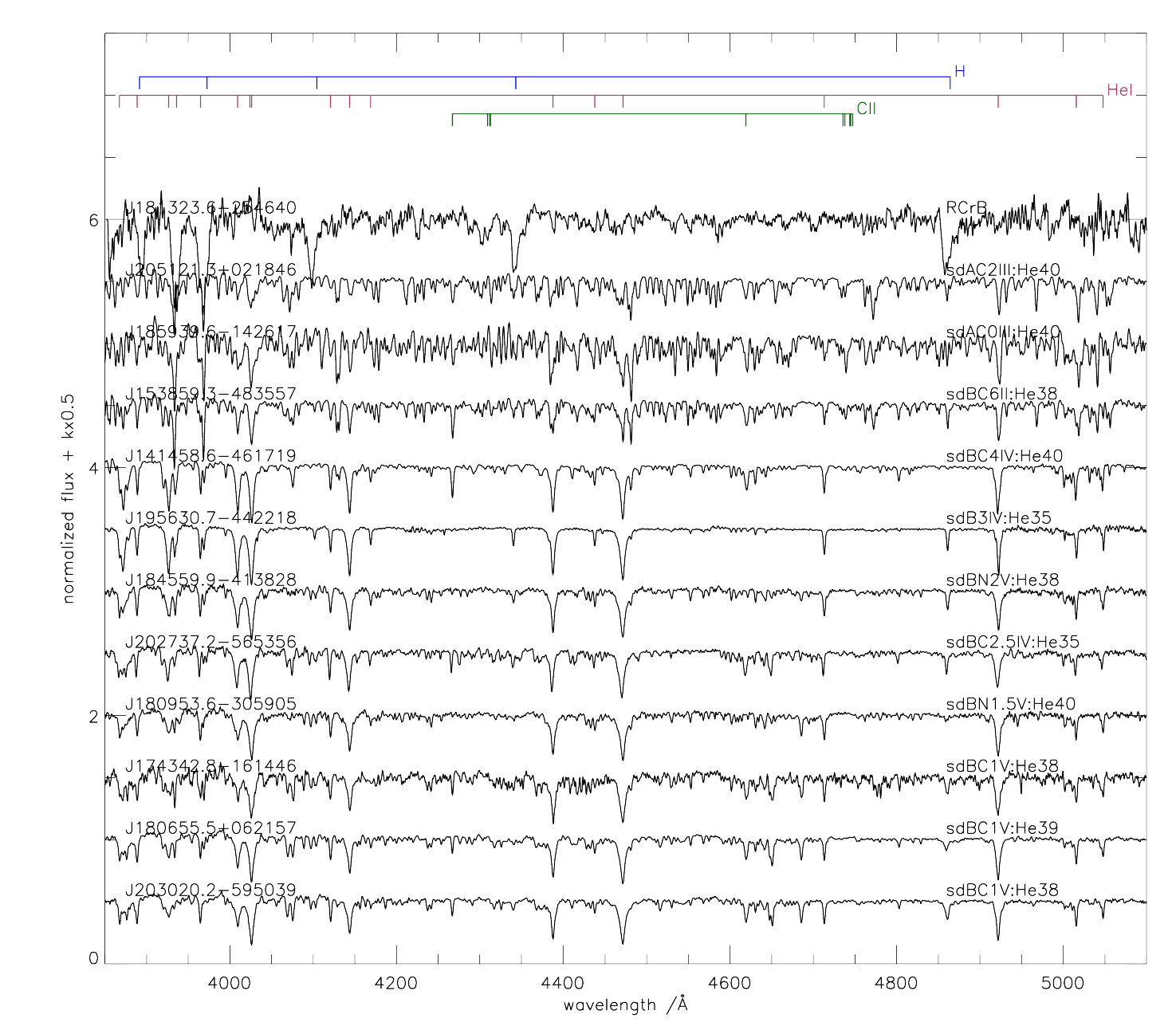}
\caption{SALT/RSS spectra of extreme helium stars from the broad \ion{He}{i} line sequence. Counting from the top, selected familiar EHes are: 2: FQ\,Aqr, and 5: V821\,Cen = HD\,124448 \citep{popper42}. } 
\label{f:ehe_hi}
\end{center}
\end{figure*}

\section{Classification}
\label{s:classes}

The entire SALT sample is described in Table\,\ref{a:classes}. {Column 1 provides a label for cross-referencing locally. Columns 2 -- 4 give positions (J2000.0), \textit{Gaia} magnitudes, and \textit{Gaia} DR3 identifiers. 
 Where appropriate, column 5 gives one additional name by which a star may be known.} 

The majority of spectra have been classified on the D13 system using direct measurements of line depths $d$ and widths $w$ as described by \citet{jeffery21a}. 
This system provides a spectral type (Sp), luminosity class (LC) and helium class (He) for each star. 
All classifications were cross-checked manually by plotting sequences in class order and correcting anomalies.
Secondary criteria, such as the ratio $d_{4713}/d_{4686}$, were used if, for example, one line in a primary criterion is weak or contaminated. 
This was particularly useful for correcting the luminosity classes and for identifying anomalies not covered by the D13 system.  
Table\,\ref{a:classes} gives the final spectral classes in alphanumeric form {(column 6)} and in numerical form including errors {(columns 9 -- 12)}. 

Other spectroscopic catalogues which include the star are indicated by abbreviation {in the final column (13).}  

Stars identified as hot pre-white dwarfs or white dwarfs do not lie within the D13 system, and have been classified accordingly ({\it e.g.} O(H), O(He), PG1159, DO, DB, DA, see Sect.\ \ref{sect:otherclass}). 
One spectrum has been identified as a probable cataclysmic variable (CV) from the presence of broad Balmer emission lines.  
D13 has few standards for very early-type subdwarfs (sdO2 -- sdO5), and was never designed for use with hydrogen-deficient giants later than B2.  
Therefore additional criteria were introduced.

\subsection{Hot sdO stars}

For stars with ${\rm Sp}<0.55$ (sdO5.5) and ${\rm He}<10$, the helium lines are too weak to establish ${\rm Sp}$. 
We define a mean Balmer line depth
\[ d_{\rm H} = \langle d_{\beta},d_{\gamma},d_{\epsilon} \rangle \]
and hence define
\[ {\rm Sp} = 0.5 - 2 (0.375 - d_{\rm H}). \]
When \ion{He}{ii}\,4541 is too weak to establish the helium class, {\it i.e.} when ${\rm Sp}<0.55$ and $d_{4541}<3\sigma_{4541}$, where $\sigma_{\lambda}$ is the standard deviation of the flux in the region of $\lambda$, 
\[ {\rm He} = 10 d_{4686} / d_{\gamma}. \]
Similarly, LC is obtained from $d_{\rm H}$ and the mean full-width half-maximum of the principal Balmer lines 
\[ w_{\rm H} = \langle w_{\beta},w_{\gamma},w_{\epsilon}  \rangle \]
as
\[ {\rm LC} = w_{\rm H} + 10.3 d_{\rm H} - 7.8, {\rm LC} \geq 8. \]
Line widths $w$ are measured by fitting either a Gaussian or parabola to the lowest 60\% of the line profile, depending on the number of available data points. 

\subsection{Extreme helium stars}

There are too few extreme helium {stars} (EHes) with ${\rm Sp}>1.2 $, (sdB2\footnote{The prefix `sd' implies that the classification is based on D13 and not necessarily that the star is a hot subdwarf; luminosity classes VI and upward imply that a star has broader lines than a main-sequence star of the same spectral class and would therefore be deemed subluminous.}), {${\ LC}\leq V$ } and ${\rm He}>35$ to establish robust classifications, but the behaviour of lines used in stars of normal composition can be used to set up spectral sequences \citep[cf.][]{gray09}.
The SALT/RSS data provide a unique panorama of EHe spectra (Figs.\,\ref{f:ehe_lo} and \ref{f:ehe_hi})\footnote{Three known EHes are not accessible to SALT: BD+10 2179 (DN Leo), LS II+33 5 (V1920 Cyg) and BD+13 3224 (V652 Her); LSS 3184 (BX Cir) has not been observed with SALT/RSS, but has been observed extensively with SALT/HRS \citep{martin19.phd}. }. 
These are distinguished by  weak or absent Balmer lines. 
The relative depths of \ion{He}{i}\,4471 and \ion{Mg}{ii}\,4482 define a spectral sequence between B0 and A0 similar to that for normal B stars; approximate parity is obtained around B6, {depending on metallicity.} 
Due to the high helium abundance, \ion{He}{i} lines are still seen at A0, but are difficult to distinguish from the overall line forest at A2. 
Stellar Ca H+K lines are visible at B2, becoming stronger than adjacent \ion{He}{i} lines by B6. However care is necessary since many EHes lie at substantial distances and the interstellar component often dominates. 
The SALT/RSS sample shows that EHes fall  into two distinct luminosity groups, as also shown by the \textit{Gaia} observations \citep{philipmonai24}.  
Here, line widths distinguish the two luminosity groups. 
All members of the luminous group show strong \ion{C}{ii} and/or \ion{C}{iii} absorption and extend from O9.5 to B8 (Fig.\,\ref{f:ehe_lo}).
We adopt a luminosity class II for this group. 
The second group show \ion{He}{i} line widths increasing from late- to early-type spectra, reflecting a nearly continuous sequence from {red} giant to {hot} subdwarf (Fig.\,\ref{f:ehe_hi}). 
For late-type EHes ${\rm Sp}>1.5$ (sdB5) and following \citet{gray09}, the ratio of \ion{Mg}{ii} and \ion{Ca}{ii} line depths has been used to define spectral types, assuming that the abundances of these species are in solar proportion. 
There is greater variety in the line spectra of this group; not all have strong carbon lines. { J180953.6-305906  (\#470\footnote{Labels \# cross-reference stars in Table A1, column 1}) and J184559.8-413827 (\#511) \citep{jeffery17b} show strong nitrogen lines.
J195630.7-442218 = EC 19529-4430 (\#581) is extremely metal poor  \citep{jeffery24}. }

\begin{figure*}
\begin{center}
\includegraphics[clip, width=0.48\linewidth]{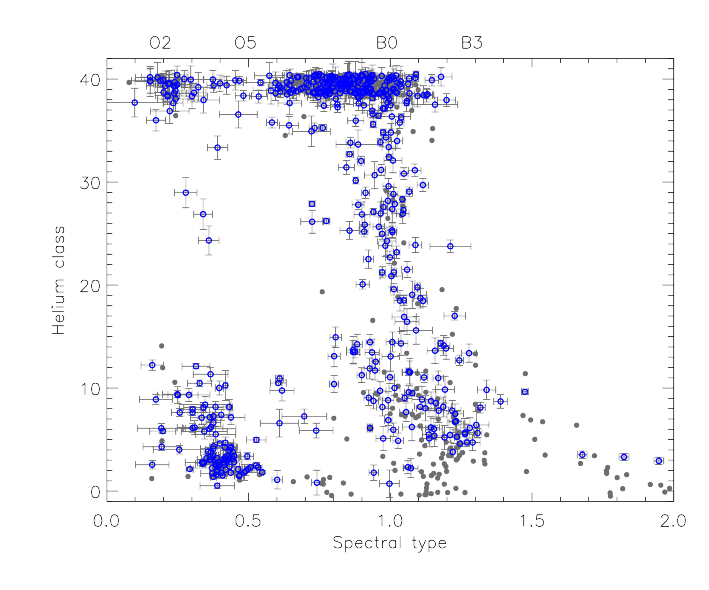}
\includegraphics[clip, width=0.48\linewidth]{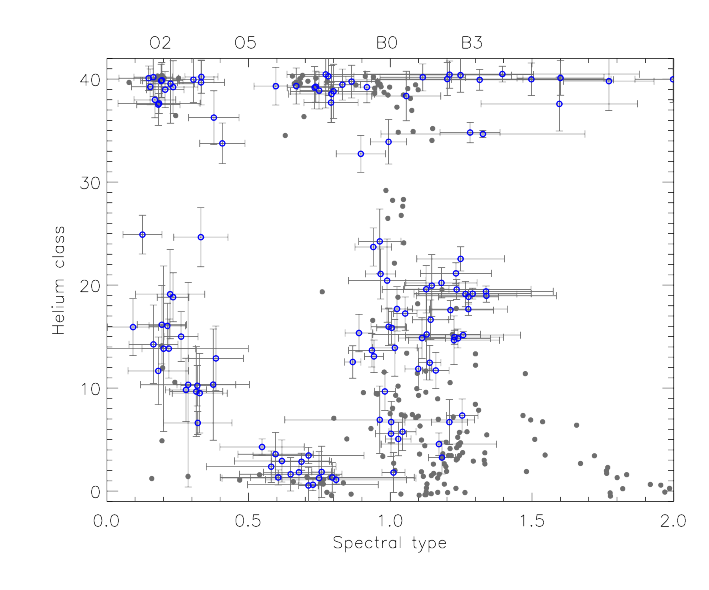}
\caption{The Sp -- He diagram for SALT/RSS observed stars with Drilling classifications (blue circles). The distribution from D13 is shown by grey dots. A uniform jitter covering $\pm$ half a division has been applied to both datasets in both axes. The left hand panel shows {515} stars for which classification errors $<0.1$ in Sp and $<1.5$ in He. The right hand panel shows the remaining {128} stars. } 
\label{f:saltclass}
\end{center}
\end{figure*}

\subsection{Other characteristics}
\label{sect:otherclass}

In addition to the principal characteristics defined by D13 (Sp, LC and He, C and N), other features have been flagged as a suffix to the spectral class. The following codes have been used to indicate the following features:
\begin{description}
\item[K:]  strong calcium K indicative of a stellar companion of late spectral type. Ca H and K lines likely due to interstellar absorption are {\it not} recorded. 
\item[K+G:] as "K" with the addition of the G-band at 4300\AA,  commonly seen in cool stars including the Sun.
\item[CO:] strong carbon and oxygen lines in an otherwise typical He-sdO star such as those identified by \citet{werner22b}.
\item[Pb:] Pb\,{\sc iv} absorption lines, notably at 4050, 4496 and 4535\AA, similar to the Pb-rich hot subdwarf EC\,22536-5304 {(\#673)}.
\item[Zr:] shows Zr\,{\sc iv} absorption lines, notably at 4198 and 4569\AA, similar to the Zr-rich hot subdwarf LS\,IV-14\,116 {(\#628)}. 
\item[rvv:] repeated spectra show radial-velocity variations with a probability of being constant $\log p<-4$.  
\item[dib:] the diffuse interstellar band at 4428\AA\ \citep{lan15}.  
\item[mag:] splitting in \ion{He}{i} and \ion{He}{ii} lines and broad unidentified absorption at 463nm as in magnetic iHe-sdOs \citep{dorsch24}.
\item[463:] as ``mag'' with Zeeman splitting not detected. 
\item[Hem:] H$\beta$ emission (not extended).
\item[em:] emission lines other than hydrogen. 
\item[NVe:] \ion{N}{v}\,4604\AA\ emission. 
\end{description}

\subsection{Other classifications}

In some cases, it is either not possible or appropriate to provide a Drilling class for the spectrum. 
Other classes appearing in Table\,\ref{a:classes} are identified as follows:
\begin{description}
\item[CV:] broad H and \ion{He}{i} absorption with some emission, comparable with a known cataclysmic variable \citep[subtype in parenthesis]{Cat.CVdownes}.
\item[O(H):] Balmer and \ion{He}{ii} 4686 absorption, the latter with central emission and also\ion{N}{v} emission. 
\item[O(He):] as O(H) but with very shallow Balmer absorption.
\item[PG1159:] strong \ion{He}{ii} 4686 and \ion{C}{iv} 4658 emission within broader absorption; other \ion{He}{ii} and \ion{C}{iv} lines may be present in absorption.
\item[DO:] broad and shallow \ion{He}{ii} lines.
\item[DB:] broad and shallow \ion{He}{i} lines.
The effective temperature and surface gravity may be provided as a suffix in the form {\it tt/gg} where  $ {\it tt}=\Teff/$kK
and ${\it gg} = \log g /{\rm cm\,s^{-2}}$. 
\item[DA:] broad Balmer lines. 
\item[RCrB:] a warm R Coronae Borealis star was included for comparison with cool EHe stars. 
\end{description}

\begin{table}
\caption{{Distribution of major {hot} subdwarf types in the SALT sample. Types defined as supersets of the Drilling classifications as shown.
The totals for HesdO and HesdB stars include the carbon-rich subclasses shown in parentheses. 
Additional stars were classified as shown; --sd-- indicates spectra too noisy to classify precisely. 
The sample is biased towards helium-rich stars and is not representative of the {hot} subdwarf population as a whole.} }
\label{t:classes}
\begin{center}  
\begin{tabular}{lrll}
\hline
Group	& $n$ & Sp & He \\		
\hline
sdB	& {74} & sdB0 - sdB3 & 01 - 14 \\
sdO	& 44 & sdO6 - sdO9.5 & 01 - 14 \\
hot sdO	& {110} & sdO1 - sdO5 & 01 - 14 \\[1mm]
iHesdOB	& {97} & sdO7 - sdB3 & 15 - 34 \\
hot iHesdO & {13} & sdO1 - sdO6 & 15 - 34 \\[1mm]
HesdB	& {38} & sdB0 - sdA0 & 35 - 40 \\
(HesdBC)	 & {(14)} & & \\
HesdO	& {189} & sdO6 - sdO9.5	& 35 - 40 \\
(HesdOC)	 & {(70)} & & \\
hot HesdO & {56} & sdO1 - sdO5 & 35 - 40 \\
(hot HesdOC) & (42) & & \\
--sd-- & {23} & \\ 
\hline
All hot subdwarfs & {644} & & \\
\hline
EHe  & {17} &  sdB0 - sdA3 & 35 - 40  \\
BHB  & 5 & sdB4 - sdA0 & 01 - 14  \\
O(H) & 4 & &  \\
O(He) & 3 & &  \\
PG1159 & 5 & &  \\
DO+DOZ & 4 & &  \\
DB+DBA & 7 & &  \\
DA+DAB+DAO & 4 & & \\
CV & 2 & & \\
RCrB, [WELS] & 2 & &   \\
\hline
All others & {53} & & \\
\hline
{Total} & {697} & & \\
\hline
\end{tabular}
\end{center}
\end{table}

\subsection{Summary}

Figure\,\ref{f:saltclass} shows the overall distribution of D13 spectral types and helium class {for the sample. 
The left-hand panel includes 515 stars for which D13 classes have been determined with high confidence. 
The right-hand panel shows 128 stars with lower confidence classifications.
A further 54 stars are not included in Fig.\,\ref{f:saltclass}.
These are either subdwarfs with an incomplete D13 classification ('-sd-' in Table\,\ref{t:classes}) or belong to classes for which a D13 classification would be inappropriate (`O(H) -- [WELS]' in Table\,\ref{t:classes}). 
 }

{A comparison of Fig.\,\ref{f:saltclass} (left)} with \citet[][Fig. 4]{jeffery21a} shows the following.  
a) A similar distribution of extremely helium-rich stars, but in greater numbers, with concentrations around sdO2 and between sdO6 and sdB1 and between He38 and He40. These are associated with the more general labels hot He-sdO, He-sdO and He-sdB. 
b) A similar distribution of intermediate helium stars (iHe-sdOB), again in greater numbers, between sdO8 and sdB1 and between He15 and He35.
c) A sizeable number of helium-poor subdwarfs of spectral type O9 or later (sdB). 
d) A large group of helium-poor subdwarfs with spectral types earlier than O6 (sdO).

{Most of the lower confidence classifications (Fig.\,\ref{f:saltclass}, right) are associated with noisy spectra. In addition, and }
for Sp $<$ sdO3 (0.3) and 10 $<$ He $<$ 30, and also for Sp $>$ sdB3 (1.1) and He $>$ 36, the errors generated by automatic classification criteria are compromised by weak \ion{He}{ii} lines (early types) and by weakening \ion{He}{i} and strengthening metal lines (late types). 
Many of the extreme helium stars (EHe) fall into this latter group. 

The Drilling classes may be approximately mapped onto the major hot subdwarf classifications sdB, sdO, HesdB, HesdO, etc. 
Table\,\ref{t:classes} shows the mapping adopted here and the numbers in each major class in the SALT survey sample. The numbers of carbon-rich HesdO and HesdB stars are also shown.
{The EHe and HesdB classes are distinguished by luminosity classes LC  II--IV and LC V--VII respectively. }

{In the final sample of 697  stars, 283 are extremely helium-rich hot subdwarfs {and 17} are extreme helium stars. {110} stars are intermediate helium-rich hot subdwarfs. The boundaries between helium-poor, intermediate and helium-rich stars set at helium classes 15 and 35 are somewhat arbitrary but have proved useful over time.  For comparison, a volume-complete sample of 305 hot subluminous stars within 500\,pc contains 30 helium-rich stars of all types \citep{dawson24}. \citet{luo24} measured carbon and nitrogen abundances for 210 helium-rich hot subdwarf stars observed in LAMOST DR7. }


\begin{figure*}
\begin{center}
\includegraphics[clip, width=0.98\linewidth]{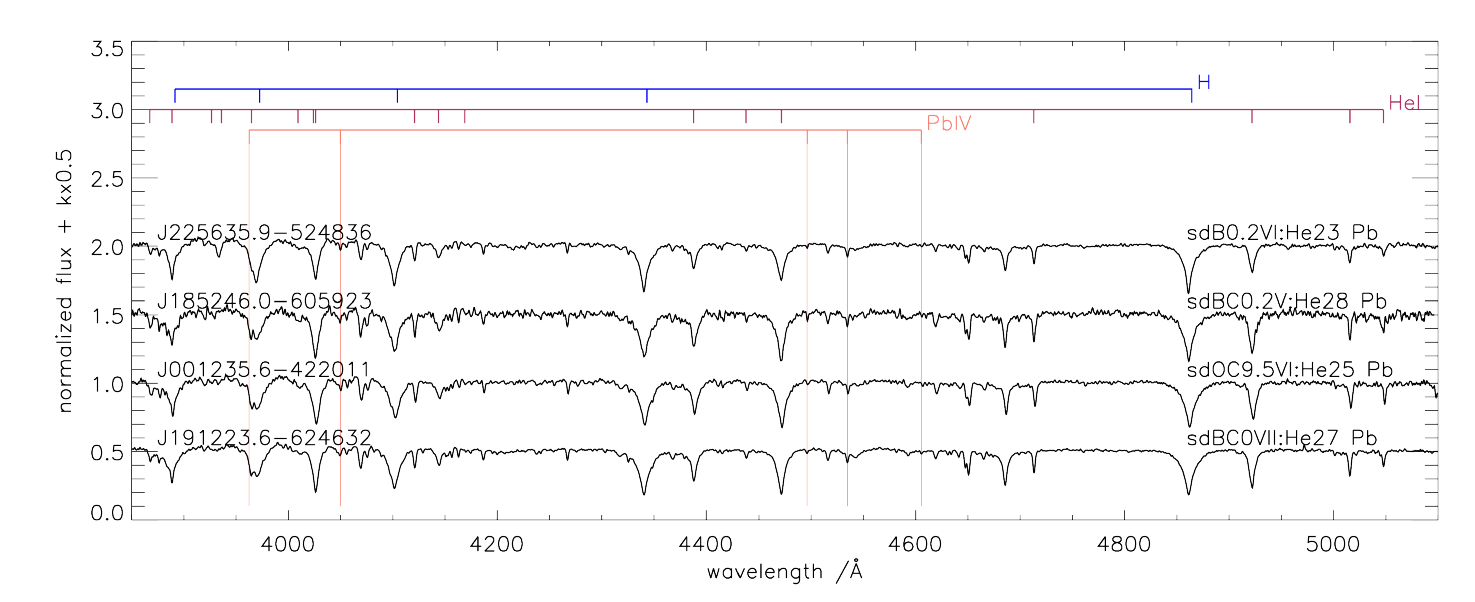}
\caption{SALT/RSS spectra of hot subdwarfs showing strong Pb{\sc iv} lines.} 
\label{f:pb}
\end{center}
\end{figure*}

\begin{figure*}
\begin{center}
\includegraphics[clip, width=0.98\linewidth]{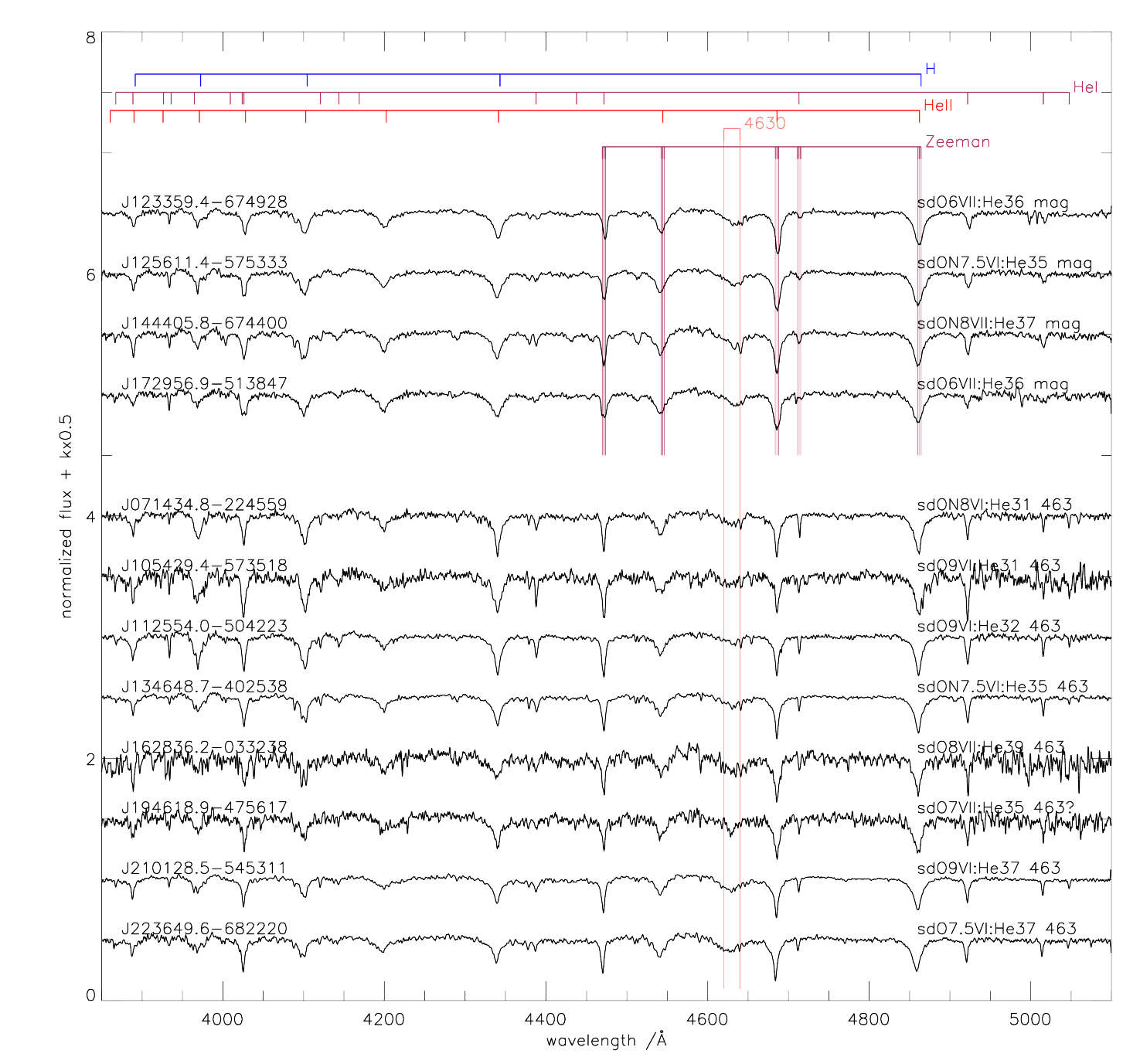}
\caption{SALT/RSS spectra of hot subdwarfs showing Zeeman-split \ion{He}{i} lines (top) and/or a broad absorption band around 4630\AA\ (bottom).} 
\label{f:mag}
\end{center}
\end{figure*}

\begin{figure*}
\begin{center}
\includegraphics[clip, width=0.98\linewidth]{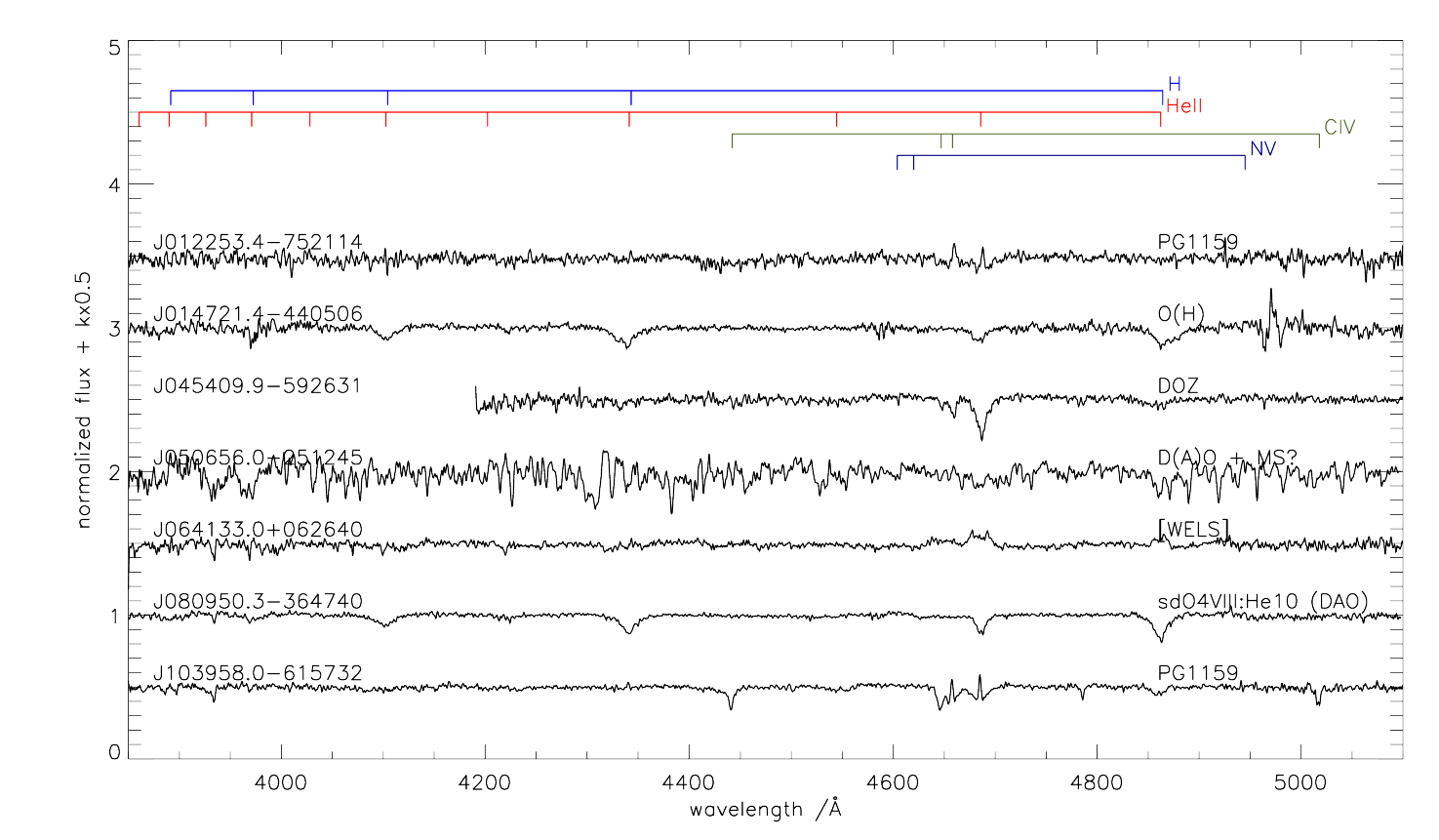}
\caption{SALT/RSS spectra of hot white dwarfs and pre-white dwarfs observed in the SALT/RSS survey additional to those reported by \citet{jeffery23a}. } 
\label{f:newhot}
\end{center}
\end{figure*}

\begin{figure*}
\begin{center}
\includegraphics[clip, width=0.98\linewidth]{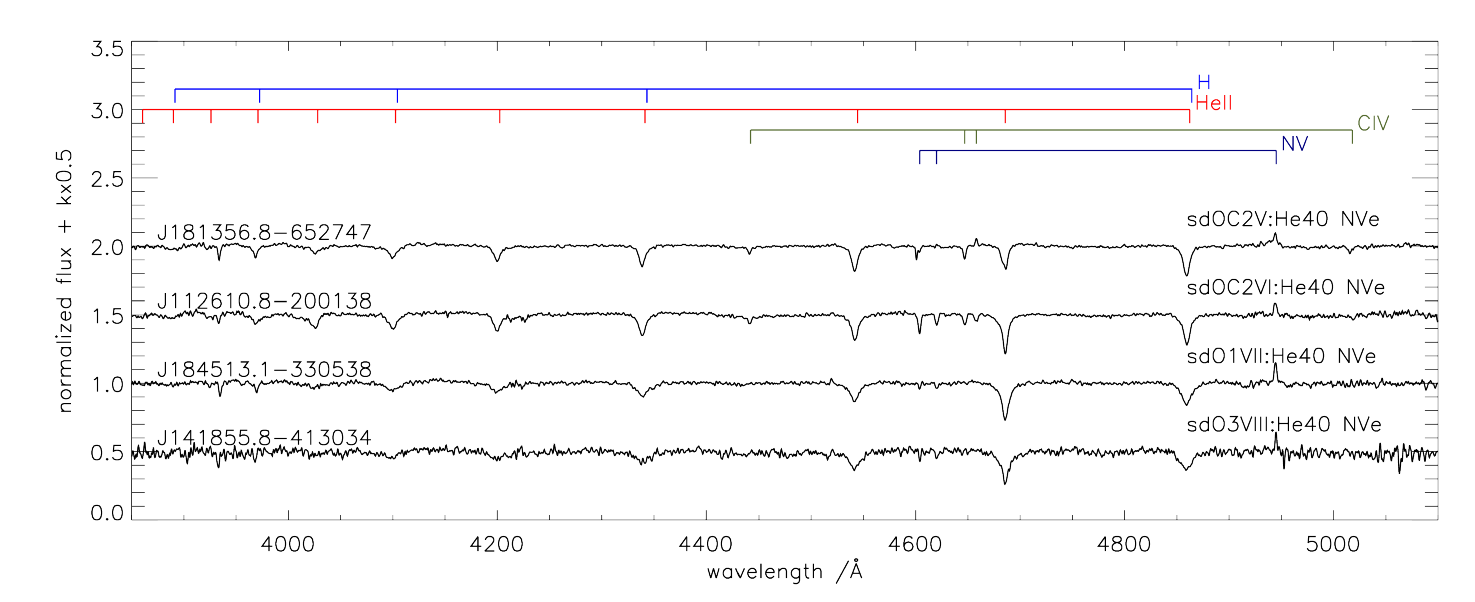}
\caption{SALT/RSS spectra of helium-rich hot subdwarfs showing\ion{N}{v} emission. } 
\label{f:hotn5}\end{center}
\end{figure*}

\begin{figure*}
\begin{center}
\includegraphics[clip, width=0.98\linewidth]{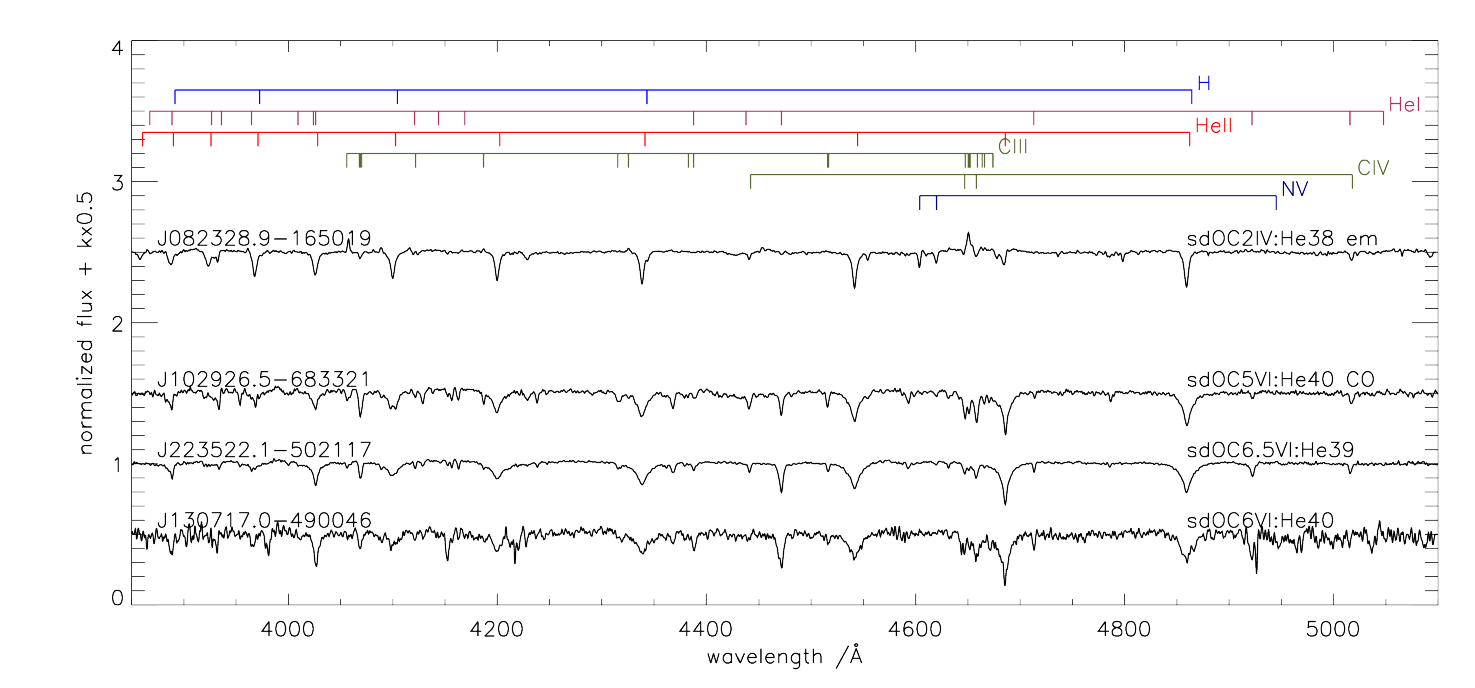}
\caption{SALT/RSS spectra of He-sdO stars showing very strong carbon lines.} 
\label{f:co}
\end{center}
\end{figure*}

\begin{figure*}
\begin{center}
\includegraphics[clip, width=0.98\linewidth]{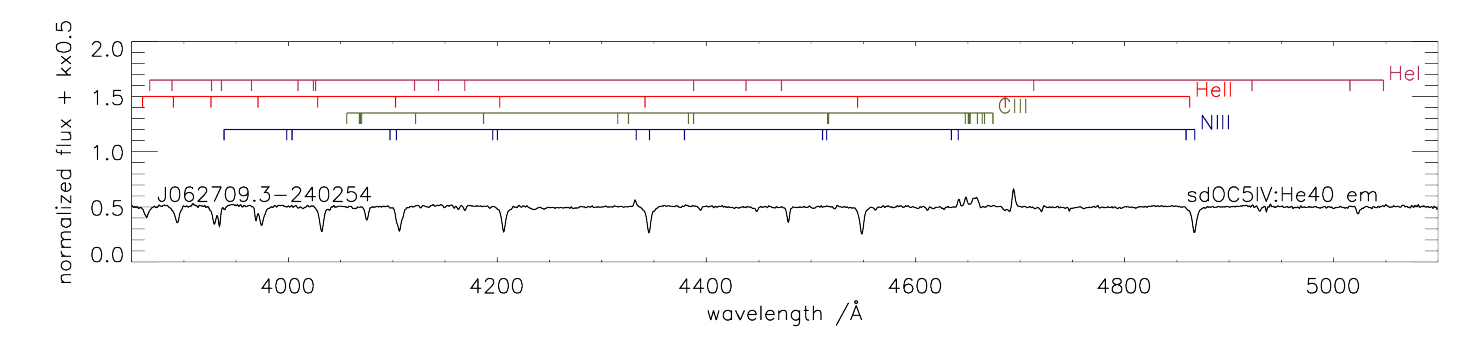}
\caption{SALT/RSS spectrum of a He-sdO star showing \ion{C}{iii}, \ion{N}{iii}, and \ion{He}{ii} 4686 emission, possibly from the irradiated atmosphere of a main-sequence secondary. } 
\label{f:wels}
\end{center}
\end{figure*}

\begin{figure*}
\begin{center}
\includegraphics[clip, width=0.98\linewidth]{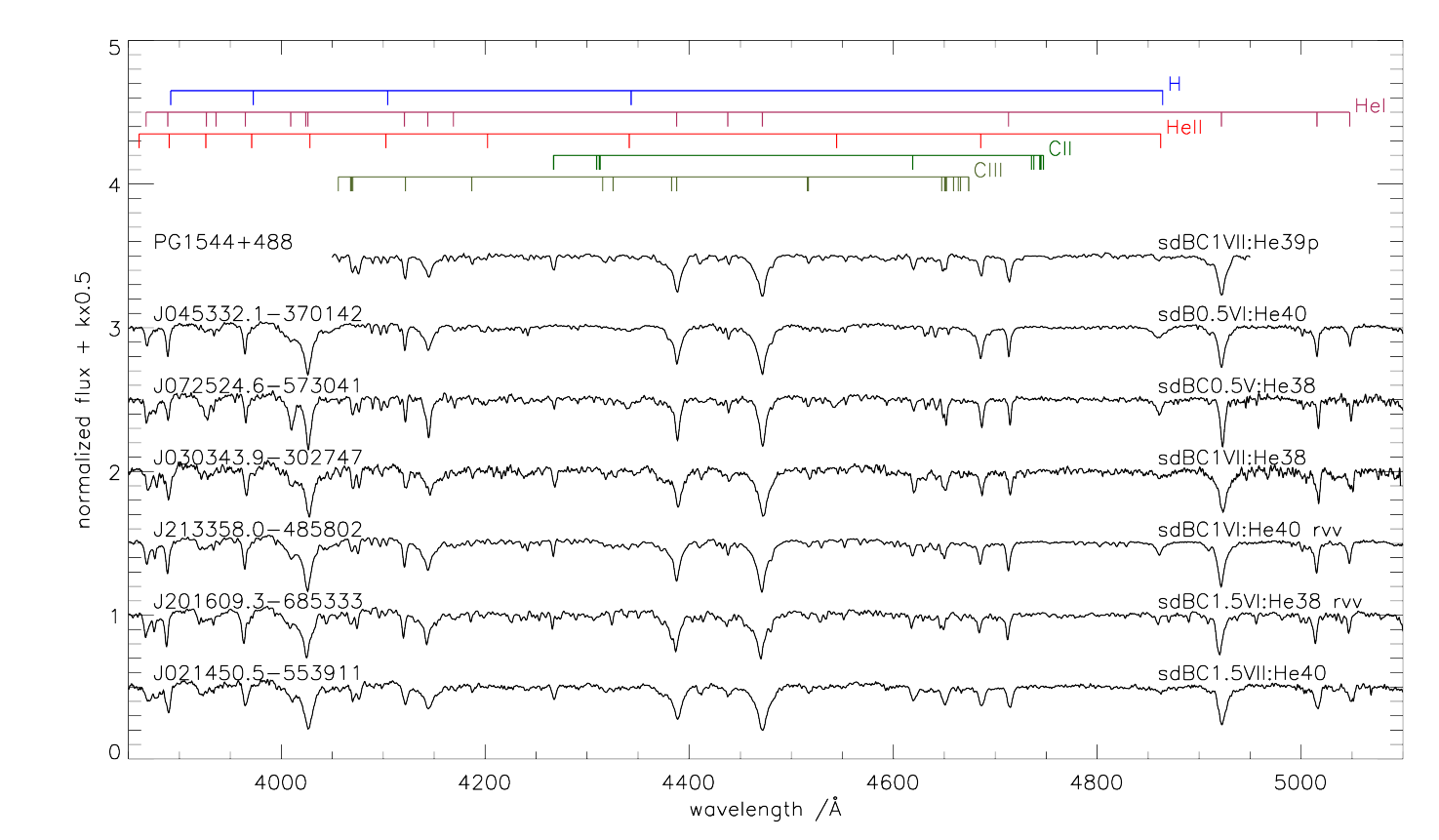}
\caption{SALT/RSS spectra of candidate double He-sdB stars. {The spectrum of PG\,1544+488 from D13 is shown for comparison} } 
\label{f:pg1544m}
\end{center}
\end{figure*}

\begin{figure*}
\begin{center}
\includegraphics[clip, width=0.98\linewidth]{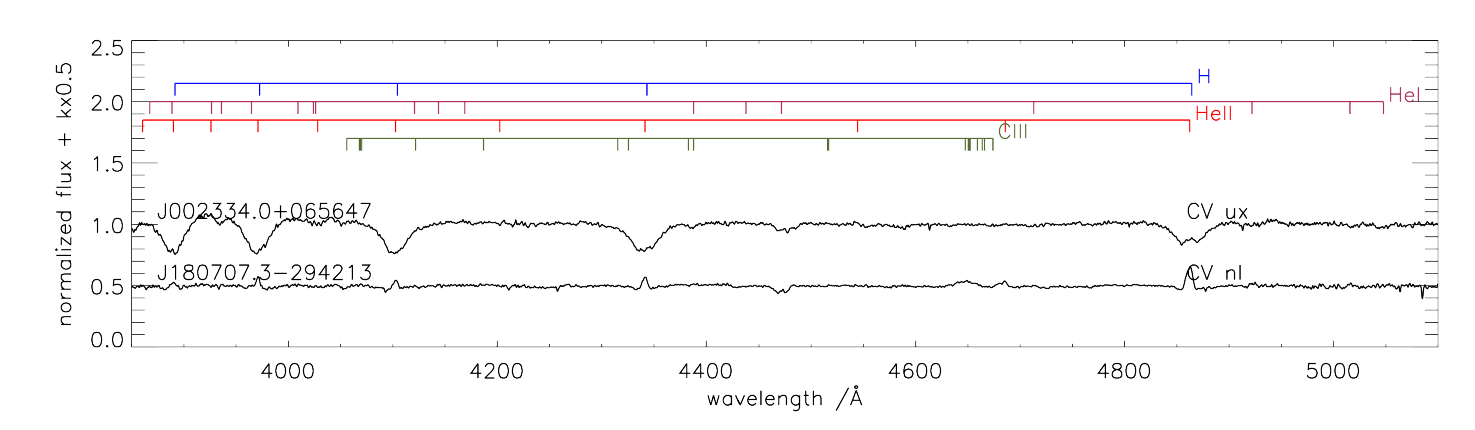}
\caption{SALT/RSS spectra of two cataclysmic variables.} 
\label{f:cv}
\end{center}
\end{figure*}

\section{Stars of interest}
\label{s:stars}

A major aim of the SALT survey is to identify stars of particular interest for further investigation. 
\citet{jeffery21a} identified several groups including:\\
a) stars at late spectral types (sdO9--sdB3) (or low \Teff\ and $g$) which might indicate links to other classes of helium-rich stars, \\
b) stars with intermediate helium classes (He10--He35) (or helium to hydrogen ratios) which might include heavy-metal stars and which confront the question of why hot subdwarfs are predominantly extremely helium-poor or helium-rich,  \\
c) stars with anomalous radial or rotational velocities which might indicate hot subdwarfs in close binary systems or otherwise high-velocity stars, and \\
d) other evolved stars of interest.\\
This paper develops some of these groups further and identifies new groups of  interest. 

\subsection{Heavy-metal hot subdwarfs}
\label{s:high-z}

The helium-rich heavy-metal hot subdwarfs comprise two  groups, those showing excess zirconium and those showing excess lead.
All form a tight group around sdO9.5 - sdB1 with zirconium-rich stars having slightly later types than lead-rich stars \citep{ostensen20}.
\ion{C}{ii}\,4267\AA\ is strong enough to justify a C-strong classification in most cases. 
Both H and \ion{He}{i} lines are strong, indicating an intermediate helium class He18 - He31. 

The zirconium-rich group includes J205738.9-142543 = LS\,IV-14 116 = V366\,Aqr {(\#628)}, Feige\,46 and J231105.1-013705 = PHL\,417 {(\#685)} \citep{naslim11,latour19b,ostensen20}.
All are small-amplitude photometric variables with periods of approximately one hour \citep{ahmad05a,latour19b,ostensen20},  probably due to \textit{g}-modes driven by a carbon-oxygen opacity bump \citep{saio19}. 
No additional zirconium-rich hot subdwarfs candidates were identified in the sample.  

The lead-rich group includes
PG 1559+048 and FBS 1749+373 \citep{naslim20},
[CW83] 0825+15 \citep{jeffery17a}, 
and HZ 44 and HD 127493 \citep{dorsch19}. 
HE 2359-2844 and HE 1256-2738 are both lead- and zirconium-rich \citep{naslim13} \footnote{Although accessible, PG 1559+048, HE 2359-2844 and HE 1256-2738 were not observed with SALT/RSS since they were already well observed.}.   

\citet{jeffery19b} identified J225635.9-524836 = EC\,22536-5304 {(\#673)} as a member of the lead-rich group from SALT spectra. \citet{dorsch21} subsequently showed it to have a metal-poor late-type companion indicating that the apparent super-abundance of lead is some 8.2 dex larger than its natural value. 
Three additional stars show spectra similar to that of EC\,22536-5304, namely: J001235.6-422011 {(\#005)}, J185246.0-6059924 {(\#517)},  and J191223.6-624632 {(\#543)}. Strong \ion{Pb}{iv} lines are visible in all cases (Fig.\,\ref{f:pb}), most notably at 4050, 4496 and 4534\AA. {The Pb{\sc iv} line at 3962\AA\ is blended with H$\epsilon$, He{\sc i} and Ca{\sc ii}. Pb{\sc iv}\,4603\AA\ is too weak to resolve in the RSS spectra. 
None of these stars} appears to have a companion visible in the spectrum. 
Five observations of J001235.6-422011 show no significant radial-velocity variation ($\sigma_{\rm rv} \approx 5.6 \kmsec$).  

\subsection{Magnetic hot subdwarfs}
\label{s:mag}

Broad flat or multiple cores of \ion{He}{i} and \ion{He}{ii} lines led \citet{dorsch24} to the discovery of three new magnetic hot subdwarfs, {J123359.4-674929 {(\#240)}, J125611.4-575333 {(\#254)} and J144405.8-674400 {(\#322)}}, as a part of this survey. 
All have strong surface multi-polar fields with strengths $\approx 200$\, kG. 
Subsequently, J172956.9-513847 {(\#433)} has been confirmed as the fourth SALT magnetic hot subdwarf. These stars have been classified with the suffix `mag'. 

All four magnetic hot subdwarfs show a broad unidentified absorption around 4630\AA\ or 463\,nm. \citet{dorsch24} reported a further six stars to show this absorption, but with no evidence for Zeeman splitting in the \ion{He}{i} lines.
These are 
J071434.8-224559 {(\#130)}, 
J134648.7-402538 {(\#281)},
J162836.2-033238 = PG 1625-034 {(\#388)},
J194618.9-475617 {(\#576)},
J210128.5-545311 = EC 20577-5504 {(\#630)}, and
J223649.6-682220 = EC 22332-6837 {(\#667)}.
To these may be added J105429.4-573518 {(\#210)} and J112554.0-504223 {(\#217)}.
These stars have been classified with the suffix `463'. 
Fig.\,\ref{f:mag} shows the mean SALT/RSS spectra for all 12 stars.

\subsection{Hot white dwarfs, pre-white dwarfs}
\label{s:hot}

\citet{jeffery23a} reported the discovery of eight new superhot white dwarfs or pre-white dwarfs from the SALT/RSS survey. 
Figure \ref{f:newhot} shows the SALT/RSS spectra of additional stars identified as white dwarfs and pre-white dwarfs.    
In order of right ascension, these include the following.

{J012253.6-752115} = RX J0122.9-7521 {(\#023)} is a ROSAT X-ray source already identified as a PG1159 star with $T_{\rm eff} \approx 180\,000 {\rm K}$, $\log g / {\rm cm\,s^{-2}} =7.5$ \citep{werner96}. 

{J014721.3-440506} {(\#034)} is likely an O(H) star but the spectrum is too poor for a useful analysis. 
\citep{vincent24} give a DO classification from a \textit{Gaia} XP spectrum, along with $T_{\rm eff} \approx 120\,000 {\rm K}$, $\log g / {\rm cm\,s^{-2}} =7.3$. 
The latter seems too high. 

{J045409.9-592631 = EC 04534-5931} {(\#080)} shows strong \ion{He}{ii\,}4686 and \ion{C}{iii/iv}\,4650,4658 and weak \ion{He}{ii}/H$\beta$ absorption and is hence a DOZ white dwarf. 
\citet{vincent24} classify this star as DO from \textit{Gaia} XP spectra, and give $T_{\rm eff} \approx 120\,000 {\rm K}$, $\log g / {\rm cm\,s^{-2}} \approx 7.8$. 

{J050656.0-051245 = EC 05048-2516} {(\#082)} (D(A)O+MS) has colours of a hot star but the spectrum shows strong contribution from a G- or K-type companion, which may produce the strong TESS signature at P=1.45 d. The cool star lines become weaker in the blue, with \ion{He}{ii} 4686 detectable. 
This is likely an analogue of UCAC2 46706450, an extremely hot white dwarf with a rapidly rotating K-type subgiant companion in a 2-day orbit \citep{werner20}. 

{J064133.0+062640} {(\#120)} shows broad emission around \ion{He}{ii} 4686, \ion{N}{iii}\,4634,4641, \ion{C}{iii/iv} 4650,4658 and H$\beta$. 
Comparable spectra are found in the central stars of NGC\,6891 and NGC\,6629  \citep[][Fig.2]{marcolino03} which have been classified as weak emission-line central stars [WELS] \citep{tylenda93}, [WC]-PG1159 stars \citep{parthasarathy98}, Of(H) stars \citep{mendez91}, {and as O(H)3 Ib(f) and [WC5/6] respectively \citep{Cat.CSPN.weidmann20}}. The [WELS] designation is weakly-defined and may indicate the presence of a binary companion \citep{miszalski11}.  
With \ion{He}{i} 4713 absent, J0641 may be hotter and more like He\,2-108. A better spectrum would be helpful.  

{J080950.0-364740} {(\#162)} is a DAO white dwarf (UCAC4 267-025892) with $T_{\rm eff} \approx 104\,000 {\rm K}$, $\log g / {\rm cm\,s^{-2}} =6.7$ and $\log {\rm He/H} \approx -1.2$ \citep{reindl23}. 

{J103958.0-615731} {(\#206)} is a PG1159 star.
\citet{pietrukowicz13} list it as UCAC4 141-049067 and OGLE-GD-WD-0001 with a period of 18.3 min. 
\citet{pietrukowicz25a} identify it as a PG1159 star and therefore 
a GW\,Vir variable from an EFOSC spectrum. 
Our better spectrum indicates $T_{\rm eff} \approx 160\,000 {\rm K}$, $\log g / {\rm cm\,s^{-2}} = 5.8$ and comparable with K1-16 (Werner, private communication).
The spectrum also shows extended H$\beta,\gamma$, \ion{He}{ii} and O[{\sc iii}] emission. 
This is unlikely to be physically associated with J1039 since the latter has galactic coordinates $l,b=288,-3$ and lies on the outskirts of the $\eta$ Carinae nebula. 

All SALT/RSS observations were obtained using a long slit; all sky background spectra were inspected for H$\beta$, H$\gamma$, \ion{He}{ii} and O[{\sc iii}] emission. 
In the few cases where these lines were unambiguously identified, the sky position coincides with a well-known \ion{H}{ii} region in the galactic plane.
In addition to J1039, examples include J064133.0+062640 {(\#120)} and J075055.56-494310 {(\#144)}.

\subsection{Hot stars showing N\,V emission} 
\label{s:nv}

Four extremely hot stars were identified with \ion{N}{v} emission at 4945\AA\ (Fig.\,\ref{f:hotn5}). 
J112610-200138 = EC 11236-1945 {(\#218)}, 
J141855.8-413034 {(\#309)}, 
J181356.9-652747 {(\#474)} 
and J184513.1-330536 {(\#510)} 
have spectral types sdO1 - sdO3 and \ion{He}{ii} line strengths consistent with no hydrogen. 
All show at least \ion{N}{v} 4604 in absorption.
Three stars show \ion{N}{v} 4620 in absorption; it is marginally in emission in J1813.

\ion{N}{v} emission lines have been observed in hydrogen-rich
central stars of planetary nebulae with optically thin nebula and \Teff>100\,000\,K \citep[{\it e.g.}\rm][]{herald11}. 
In such cases, all three optical lines (4604\AA, 4620\AA, and 4945\AA) are in emission. 
The NVe stars in this survey appear to represent a new class of helium-rich hot subdwarf, possibly nitrogen-strong analogues of the carbon-strong sequence, with a weak radiatively-driven wind. 

\subsection{CO-rich He-sdO stars}
\label{s:co}
A new class of hot subdwarfs with very high surface concentrations of carbon and oxygen was identified by \citet{werner22b}, with the two prototypes PG\,1654+322 and PG\,1528+025.
A third member was announced by \citet{werner25}. 
It is identified in the SALT sample as J102926.5-683321 {(\#201,} sdOC5VI:He40 CO; Fig.\,\ref{f:co}). 
A search for similar spectra yielded the somewhat noisy J130717.0-490046 {(\#262,} sdOC6.5VI:He39) and the less C-strong J223522.1-502117 {(\#665,} sdOC6VI:He40). 
Unfortunately, the definitive \ion{O}{iii} lines at 3700-3800\AA\ and 5593\AA\ lie outside the range observed with SALT/RSS, and the \ion{O}{iii} and \ion{O}{iv} within this range are too weak to resolve at the dispersion observed. 
Both were classified with the suffix `CO?'. 

Figure \ref{f:co} also shows the spectrum of J082328.9-165019 {(\#166,} sdOC2IV:He38 em). The \ion{He}{ii} line profiles imply an early spectral type, with \ion{He}{ii} 4686 being filled in by emission, and \ion{C}{iii} 4057, 4187, 4382, and 4651 also in emission. 
\ion{C}{iv} and \ion{N}{v} lines are in absorption. 
Although several He-sdO stars show some \ion{C}{iv} emission, this spectrum appears to be unique. 

\subsection{A He-sdO+MS binary?}
\label{s:wels}  

{J062709.2-240254} {(\#117)} is unique within the SALT sample, showing significant emission at \ion{C}{iii}, \ion{N}{iii} and \ion{He}{ii} 4686. 
Otherwise, \ion{He}{ii} and \ion{He}{i} absorption lines support the  sdOC5IV:He40 classification (Fig.\,\ref{f:wels}). 
There is no evidence for a planetary nebula, so a weak emission-line central star ([WELS]) classification might be premature \citep{tylenda93}. 
Comparison with the central star of NGC\,6326 suggests emission could arise from the irradiated atmosphere of a main-sequence companion \citep{miszalski11} which might be indicated by the flat spectral energy distribution (Dorsch et al. in prep.) in the optical.
Otherwise emission could arise in the wind from the low-gravity hot star. 
Two spectra obtained 10 months apart give the same rotational and large radial velocities (ibid.).  
From its kinematics, Philip Monai et al. (2025, MNRAS submitted) identify this to be a Galactic halo system.

\subsection{Double He-sdB candidates}
\label{s:pg1544}

The prototype He-sdB star, PG\,1544+488 \citep{heber88} (sdBC1VII:He39p: D13) turned out to be a double-lined spectroscopic binary containing {\it two} helium-rich sdB stars with an orbital period of $\approx 12$ h \citep{ahmad04a,sener14}. 
\citet{jeffery21a} identified three stars with spectra similar to PG\,1544+488: J045332.1-370142 = EC\,04517-3706 {(\#079)},  J201609.3-685333 = EC\,20111-6902 {(\#595)} and  J213358.0-485802 = EC\,21306-4911 {(\#639)}.
Indications for a double He-sdB include variable line widths, variable radial velocities, periodic doubling of the line profiles and/or a high projected rotation velocity in a model fit. 
Repeat observations of all three candidates were obtained with SALT/RSS. 
EC\,21306-4911 shows significant radial velocity variations over ten epochs, but a unique period has so far proved elusive.
EC\,04517-3706 and EC\,20111-6902 were observed on 4 and 14 epochs, respectively. On the basis of the standard deviations and mean errors in the radial velocities, radial velocity variations were not deemed to have been detected (see \S\,\ref{s:rvv}). 

J030343.9-302747 = HE\,0301-3039 {(\#052,} sdBC1VII:He38) was proposed as a double He-sdO binary by  \citet{lisker04}. 
Our SALT/RSS spectrum resembles PG\,1544+488 but, so far, variability and/or a double-lined signature have not been confirmed (Fig.\,\ref{f:pg1544m}).

Additional candidates were identified in  the second part of the survey (Fig.\,\ref{f:pg1544m}). 
J021450.6-553913 {(\#041,} sdBC1.5VII:He40) is amongst the broadest lined stars with an appropriate spectral type in our sample, with $ \langle v_{\rm rot} \sin i \rangle \approx 200$\kmsec.
It shows a strong spectral similarity to PG 1544+488. 
Observations over six epochs covering 13 months show radial-velocity range of nearly 50\,\kmsec\ as well a probability of being constant $\log p = -5.2$ (\S\,\ref{s:rvv}).
The apparent line width is variable ($ v_{\rm rot} \sin i \approx 127 - 250$\,\kmsec); if caused by the reflex acceleration of two similar components, higher-resolution spectroscopy will be necessary to separate their contributions. 

J072524.6-573041 {(\#136,} sdBC0.5V:He38) also shows broad lines ($v_{\rm rot} \sin i \approx 206$\,\kmsec, but only one epoch has been observed. 

\subsection{Rapidly rotating stars}

Other helium-rich stars with large apparent $v_{\rm rot} \sin i$ include the magnetic hot subdwarfs (\S\,\ref{s:mag}), where Zeeman splitting provides additional line broadening, and a small group of hot He-sdO stars: J110647.7-572057 {(\#215)}, J181356.9-652727 {(\#474)}, J201318.8-120119 {(\#593)}, J145538.5-384304 {(\#327)} and J145807.6-381423  {(\#329)}  (spectral types sdO1 - sdO3). These may simply be "fast" rotators with $v_{\rm rot} \sin i \approx 40 - 80$\kmsec. 

Despite its classification, {J012801.4-524557 = JL\,246} {(\#024,} sdO4VII:He34 Hem) is a relatively helium-poor hot subdwarf with broad Balmer lines and a central H$\beta$ emission which cannot be explained by non-LTE effects or a stellar wind. 
It is one of a handful of stars with a projected rotation velocity over 200\,\kmsec. 
Two other hot fast rotators include {J062011.5-583904} {(\#114,}  sdO1VII:He01) and {J055805.5-521717} {(\#102,} sdO4VI:He06); neither show H$\beta$ emission. 

{
The substantially cooler but equally rapid rotators, {J203540.6-074029} {(\#613,} sdB1.5IV:He14), {J085852.6-131422} {(\#178,} sdB3V:He05) and {J074751.5-253404} {(\#143,} sdB8VII:He03) appear spectroscopically as main-sequence B or sdA in case of the latter. 
{J074751.5-253404} is reported to be a single-lined spectroscopic binary \citep[$P=0.885$\,d;][]{gaia23.dr3.mult}, meaning that some of the line broadening might be due to spectral smearing. 
This star is similar to the non-rotating {J094651.4-380321} {(\#188,} sdB7V:He04), and both stars are similar to the pre-helium white dwarf SDSS J160429.12+100002.2 \citep{irrgang21}. 
Stars spectroscopically similar to {J085852.6-131422} include {J025539.3-511132} {(\#051,} sdB2I:He14), {J063604.7-274008} {(\#119,} sdB2.5II:He18), and {J141420.5-365341} {(\#304,} sdB1I:He12). Although their spectra resemble those of main-sequence or slightly evolved B-type stars, their luminosities lie between the main sequence and the horizontal branch. 
This indicates that they may be stripped stars, analogous to the stripped star + WD system HZ\,22 = UX\,CVn \citep{Young1982,Shimansky2002}, a possibility that will be examined in a future study. 
Another relatively cool star, {J094651.4-380321} {(\#188,} sdB7V:He04), may be classified as BHB {(blue horizontal branch)}, and a corresponding Galactic halo membership could explain its high radial velocity.
}

\subsection{Extreme helium stars}
\label{s:ehe}

An original goal of the SALT survey was to identify new extreme helium stars and explore connections between cooler hydrogen-deficient stars, including the RCrB {(R Coronae Borealis)} variables and the dustless hydrogen-deficient carbon giants (HdC stars), and the helium-rich hot subdwarfs, {\it i.e.} the He-sdB  and He-sdO stars.  

Two {\it bona fide} EHe stars discovered with SALT have already been reported: J184559.8-413827 {(\#511)} and J195630.7-44218 = EC\,19529-4430 {(\#581) }\citep{jeffery17a,jeffery24}. 
The former is spectroscopically similar to the pulsating EHe star V652\,Her \citep{jeffery99b}.

Other recent discoveries of cool EHe stars include J181135.6+015432 = A980 = TYC 435-2248-1, J182448.9-221429 = A208 and J183357.0+052917 = A798 \citep{tisserand22}.
J182448.9-22145 {(\#482,} sdBC2.5II:He40) and J183357.0+052917 {(\#492,} sdBC2:He40) have similar spectra and belong to the narrow \ion{He}{i} line  sequence (Fig.\,\ref{f:ehe_lo}). They lie between J065446.2-104832 (LSS\,99) and J174425.5-193753 (LSS\,4357), implying $T_{\rm eff}\approx 16\,000$\,K and $\log g\approx 2.0$ \citep{jeffery98a}. 
J181135.6+015432 {(\#471,} sdBC8II:He40) is the coolest star in the narrow \ion{He}{i} line  sequence.  An abundance analysis revealed $\approx 1$ dex  excess of \textit{s}-process elements germanium, yttrium, strontium, zirconium and barium \citep{saini25}.  
J181323.57-254640.86 = V3795 Sgr {(\#472)} is a warm RCrB star \citep{tisserand20}. 
SALT/RSS spectra for all four stars have been obtained to illustrate the cool end of the spectral progression from He-sdB through EHe stars to RCrB (Figs.\,\ref{f:ehe_lo},\ref{f:ehe_hi}).

Additional EHe stars discovered with SALT include the following.
J202737.2-565356 = EC 20236-5703 {(\#607,} sdBC2.5IV:He35) was identified as He-rich in the Edinburgh-Cape survey \citep{Cat.EC3} and as an EHe star by \citet{jeffery21a}. Like EC\,19529-4430, V652\,Her and J184559.9-413828, it has relatively strong Balmer lines. Also like J184559.9-413828, \ion{He}{ii}\,4686 is just detectable. \ion{C}{ii} lines are prominent.   
J180953.6-305906 = [DSH99] 456-3 {(\#470,} sdBN1.5:He40) is a nitrogen-rich EHe of slightly earlier spectral type than V652\,Her and J184559.8-413827. It was classified `He+' by \citet{dufton99}, meaning `target may have an enhanced helium spectrum'. 
Being similar to V652\,Her and BX\,Cir, J1809 is a potential large-amplitude pulsator but, since $V=13.99$ and the W\,UMa-type variable OGLE BLG-ECL-33979 ($V=18.3$) is only 11 arcsec away \citep{Cat.OGLE.blg}, it seems unlikely that it would not have been identified as such already.  

Figure\,\ref{f:ehe_hi} shows a continuous sequence of broad-lined EHe spectra up to spectral type sdB1. This sequence continues to earlier spectral types and in increasing numbers, but with spectra that are more accurately labelled as hot subdwarfs. 
The hottest star shown is J203020.2-595039 = BPS\,CS\,22940-0009 {(\#610}) which was identified as a nitrogen-rich He-sdB connecting hotter He-rich subdwarfs with the EHes \citep{naslim10,snowdon22}. 
J174342.9-161446 {(\#447,} sdBC1:He38) is a new discovery and is a near twin of J180655.5+062157 = LS IV +06 2 {(\#466,} sdBC1:He39) and J2030. 

With the northern EHes and BX\,Cir, the total number of {\it bona fide} EHe stars now stands at 27, of which six have been identified with SALT. 

\subsection{Cataclysmic Variables}
\label{s:cv}

J002334.0+065647 = PB 5919 {(\#007,} CV of UX UMa type) shows broad Balmer lines with central reversals (Fig.\,\ref{f:cv}).
A bright cool component is present in the spectral energy distribution which is unlikely to be stellar.
The spectrum is similar to those of IX Vel \citep{kara23} and RW Sex \citep{hernandez17}, both nova-like cataclysmic variables of the UX UMa type \citep{Cat.CV7}. 
The TESS light curve (sector 70) is noisy but shows a strong signal with period 104.6 min and amplitude 0.04\%; first and second harmonics are also present; this is likely to be the orbital period.  
Follow-up spectroscopy was carried out on 2024 June 24, July 21 and 23; SALT/RSS was used to obtain 20 consecutive 70\,s exposures on each visit. No shifts in line position were detected within or between runs, but the line-core profiles changed from run to run.

{J180707.3-294213 = [DSH99] 456-21} {(\#467,} CV nl) shows weak central Balmer and \ion{He}{i} emission within a broader absorption component (Fig.\,\ref{f:cv}). \ion{He}{ii} 4686 and \ion{C}{iii} 4650 show weak emission. By comparison with the spectral atlas of \citet[Figure 1]{szkody11}, the nearest analogues are the nova-like  systems SDSS J123255.10+222209.4 = PG 1230+226 and SDSS J151915.86+064529.1, both of which show variable emission \citep{Cat.CVinight23}.
In contrast, \citet{dufton99} reported no \ion{H}{i}, \ion{He}{i}, \ion{Ca}{ii}, but \ion{Na}{i} present.  
\citet{barlow22} report a \textit{Gaia} variability index 0.049 on a scale $0 -1$. 

\label{s:misc}

\begin{table} 
\begin{center} 
\caption{Mean and standard deviation radial velocities for stars with four or more measurements or a probability of being constant $p<-10^{3}$. 
The stars are ordered by decreasing $\log p$. } 
\label{t:rvcands} 
\setlength{\tabcolsep}{2pt}  
\begin{tabular}{rllrrrr}  
\hline 
\# & Star & & $\langle{v_{\rm rad}}\rangle$ & $\sigma_{v_{\rm rad}}$ & $\log p$ & $n$ \\
\hline  
661 & J222122.6+052458 & PG 2218+051 & 19.4 & 1.6 & -0.0 & 6\\
031 & J014307.5-383316 & SB 705 & -8.0 & 2.7 & -0.0 & 5\\
597 & J201815.4-383359 & EC 20149-3843 & 25.4 & 5.7 & -0.1 & 6\\
005 & J001235.6-422011 &  & 52.9 & 7.1 & -0.2 & 5\\
322 & J144405.8-674401 &  & 2.5 & 6.1 & -0.2 & 4\\[1mm]
254 & J125611.4-575333 &  & 13.9 & 12.3 & -0.3 & 4\\
240 & J123359.4-674928 &  & 86.1 & 8.2 & -0.4 & 4\\
079 & J045332.1-370142 & EC 04517-3706 & 3.5 & 7.1 & -0.4 & 4\\
633 & J211111.5-480256 & EC 21077-4815$^a$ & 64.5 & 8.3 & -0.4 & 8\\
656 & J221656.0-643150 & BPS CS 22956-0094$^a$ & -27.1 & 8.5 & -0.5 & 11\\[1mm]
173 & J084528.8-121409 &  & 107.5 & 9.0 & -0.7 & 4\\
592 & J201118.7-315606 & EC 20081-3205 & -66.4 & 13.9 & -1.0 & 5\\
014 & J004905.1-542438 & EC 00468-5440 & 64.1 & 13.2 & -1.2 & 8\\
595 & J201609.3-685333 & EC 20111-6902 & -85.2 & 12.1 & -1.5 & 14\\
630 & J210128.5-545311 & EC 20577-5504 & -44.4 & 13.8 & -1.7 & 12\\[1mm]
090 & J052612.3-285824 & EC 05242-2900 & 26.4 & 13.0 & -1.7 & 4\\
008 & J002341.7-532017 & JL 183 & 87.1 & 33.6 & -5.7 & 3\\
204 & J103413.5-524919 &  & 20.3 & 29.7 & -7.4 & 3\\
586 & J200439.2-424627 &  & 10.3 & 35.2 & -8.3 & 2\\
495 & J183659.5-462201 &  & -17.7 & 87.1 & -50.2 & 2\\[1mm]
639 & J213358.0-485802 & EC 21306-4911 & -0.2 & 46.1 & -58.8 & 10\\
087 & J051756.5-304749 & EC 05160-3050$^{a,b}$ & 138.3 & 131.3 & -227.0 & 4\\
\hline  
\end{tabular}  
\parbox{6cm}{
$a$: TESS periodic variable: \citep{snowdon25}\\
$b$: Ton\,S\,415: $P = 86.4$\,min \citep{snowdon23a}}
\end{center} 
\end{table}  

\subsection{Photometric and radial-velocity variable stars}
\label{s:rvv}

J051756.5-304749 = EC\,05160-3050 = Ton\,S\,415 {(\#087)} was identified as a large-amplitude radial-velocity variable and hence as a short-period iHe-sdO+WD binary from SALT/RSS data \citep{snowdon23b}. 
It also shows a photometric variation at the orbital period. 

An early cross-match for photometric variables amongst SALT/RSS stars suggested 16 candidates, 10 of which turned out to be false positives \citep{snowdon25}.  
Four stars with TESS periods showed no evidence for any radial-velocity variation, including 
J221656.0-643150 = BPS\,CS\,22956-0094  {(\#656)}, 
J041319.4-134102 = EC\,04110-1348 {(\#064)},
J211111.5-480256 = EC\,21077-4815  {(\#633)},
and J014307.5-383316 = SB\,705  {(\#031)}.
J123406.0-344532 = TYC 7242--541--1 {(\#241)}, with a TESS period of 0.396\,d,  was only observed once with SALT/RSS. 
A further 137 extreme and intermediate helium stars observed with SALT/RSS show no photometric variability in TESS \citep{snowdon25}. 

\citet{krzesinski22} report J191849.6-310441 {(\#555} sdB3VII:He06) to be a \textit{g}-mode pulsator with its largest amplitude pulsation at a period of 55.5\,min and an amplitude of 0.0045 mag. Two SALT/RSS measurements show no significant difference in velocity. 

Radial velocities have been measured independently for all SALT/RSS spectra by both 
Philip Monai et al. (2025, MNRAS submitted) and Dorsch et al. (2025, in prep).
Repeat observations were made for several potential radial-velocity and spectrum variables and also when a star was too faint to observe within a single visit.
The likelihood of a star's radial velocity being constant \citep[$p$:][]{maxted01} can be estimated by comparing the velocity standard deviation with the measurement error. 
Table\,\ref{t:rvcands} summarizes the radial velocity statistics for all stars with four or more velocity measurements or for which $\log p<-3$ {(column 6)}. 
The latter have been classified 'rvv' in Table\,\ref{a:classes}. 
{Other columns in Table \,\ref{t:rvcands} include the mean ($v_{\rm rad}$: column 4), standard deviation ($\sigma$: column 5) and number of observations ($n$: column 7).}

The cases of J045332.1-370142 = EC 04517-3706, J201609.3-685333 = EC 20111-6902   and J213358.0-485802 = EC 21306-4911 
were discussed in  (\S\,\ref{s:pg1544}). 
Only EC\,21306-4911 is emphatically variable. 

Only two observations are available for each of {J183659.5-462201} {(\#495)} and {J200439.2-424627} {(\#586)}.   
Shifts of 175 and 70 \kmsec\ were detected between pairs of otherwise excellent spectra obtained approximately 60 and 40 days apart, respectively.  
Both are hot sdO stars and would be excellent candidates for follow-up spectroscopy. 
{J103413.5-524919} {(\#204,} sdB2.5VI:He08) is a typical sdB star showing a velocity range of 66 \kmsec\ over 3 epochs. 
{(J002341.7-532017 = JL\,183} {(\#008,} sdO9.5VII:He24) shows a velocity range of 81 \kmsec over 3 epochs.  
Both also warrant follow-up spectroscopy. 

It is noted that {the long-term stability of SALT/RSS is not ideal for} radial-velocity measurements except in the case of repeat observations of large-amplitude variables over short intervals of time \citep[{\it e.g.}\rm][]{snowdon23b}. 

\begin{table}
    \caption{Survey stars with very high radial velocities. }
    \label{t:high_rv}
    \setlength{\tabcolsep}{2pt}  
    \begin{center}  
       \begin{tabular}{rllrlcc}
       \hline
       \# & Star &    & \multicolumn{2}{c}{$v_{\rm rad}$/\kmsec} & $\sigma_{v_{\rm rad}}$ & $n$ \\
       \hline
563 & J193046.0-305000 & & $-$328.4 & $\pm$5.2 & -- & 1 \\
604 & J202418.0-070835 & & $-$303.2 & $\pm$5.9 & -- & 1 \\
437 & J173225.2-614205 & & 229.9 & $\pm$8.0 & -- & 1 \\
046 & J023326.1-591231 & EC 02320-5925 & 233.9 & $\pm$5.9& 5.2 & 3 \\
065 & J041400.2-315443 & EC 04120-3202 & 240.7 & $\pm$9.3& -- & 1 \\[1mm]
272 & J132307.8-480542 & & 251.7 & $\pm$15.7& 4.3 & 2 \\
380 & J161517.8-212604 & & 276.2 & $\pm$6.5& -- & 1 \\
051 & J025539.3-511132 & & 279.5 & $\pm$5.3& -- & 1 \\
517 & J185246.0-605924 & & 327.8 & $\pm$5.4& -- & 1 \\
311 & J142100.1-362522 & & 336.3 & $\pm$6.1& -- & 1 \\[1mm]
188 & J094651.4-380321 & & 364.4 & $\pm$6.6& 7.6 & 3 \\
472 & J181323.6-254641 & & 396.8 & $\pm$15.9& -- & 1 \\
117 & J062709.3-240255 & & 459.0 & $\pm$5.3& 0.2 & 2 \\
\hline
        \end{tabular} 
    \end{center}
\end{table}

\subsection{High radial velocity}
\label{s:hivrad}

Table\,\ref{t:high_rv} identifies all sample stars with a radial velocity $|v_{\rm rad}|> 220$\kmsec. 
These are interesting because they could include potential runaway stars, halo stars, or otherwise kinematically energetic stars, such as binaries.
{The standard deviation $\sigma$ is shown for stars with more than $n=1$ observations. }

To identify the latter, additional observations are required for all stars in the table. 

From a subsample of 584 stars including all stars in Table\,\ref{t:high_rv}, Philip Monai et al. (2025, MNRAS submitted) find no  unbound stars and only 48 halo stars, corresponding to $<10\%$ of the sample. 
The latter are distributed amongst most sample spectral classes except the hydrogen-rich sdB stars.

\section{Conclusion}

The SALT survey of helium-rich {hot} subdwarfs aims to characterize the properties of a substantial fraction of {such stars} in the southern hemisphere, to establish the existence and sizes of subgroups within that sample, and to provide evidence with which to explore connections between these subgroups and other classes of evolved star.

This paper presents the final sample of $697$ stars, their positions, \textit{Gaia} DR3 magnitudes and identifiers, their reduced SALT/RSS spectra, and their spectral classifications. 
It now represents the largest homogeneous sample of both "normal" He-sdOs and "luminous" or "hot" He-sdOs, and will enable a more systematic study of WD mergers and the post-HeMS evolution of He-sdOs.
The spectra will be used immediately for papers on (a) an unsupervised spectral classification and kinematic analysis (Philip Monai et al. 2025, MNRAS submitted) and (b) atmospheric parameters and evolution inferences
(Dorsch et al., in prep.). 

The identification of two distinct sequences of EHe stars based on Gaia distances and luminosities by \citet{philipmonai24} is strongly supported by the spectral classifications, with the lower-luminosity sequence connected directly to the "normal" He-sdOs on the He-MS. 
More puzzling is the scarcity of hot luminous EHe stars; these are needed to prove a connection with O(He) stars or other hot or luminous He-sdO stars. 
A demonstration of how either sequence connects with the cooler RCrB stars is beyond the scope of the current sample, but there is evidence for both. 

The paper also identifies a number of `exotic' stars with unusual spectral characteristics. 
Some of these, such as the \ion{N}{v} emission-line stars, have not been seen before in similar hot subdwarf surveys and will require detailed analysis. 
Several, including the new double-subdwarf candidates, the Pb stars and several apparently rapidly rotating B-type stars, merit further follow-up observation with either better time resolution,  better spectral resolution or both.

\section*{Acknowledgments}

The Armagh Observatory and Planetarium is funded by direct grant form the Northern Ireland Department for Communities. That funding has enabled the Armagh Observatory and Planetarium to be a member of the United Kingdom SALT consortium (UKSC), which is a shareholder in the Southern Africa Large Telescope (SALT).  
All of the observations reported in this paper were obtained with SALT following  generous awards of telescope time from the UKSC and South African SALT Time Allocation Committees. 

MD was supported by the Deutsches Zentrum für Luft- und Raumfahrt (DLR) through grant 50-OR-2304. 
For some of the time covered by the SALT survey; CSJ was part funded by the UK Science and Technology Facilities Council (STFC) grant no. ST/M000834/1.

This research has made use of the SIMBAD database,
operated at CDS, Strasbourg, France 

This paper makes use of services or code that have been provided by the AAO Data Central Science Platform (datacentral.org.au).

\section*{Data Availability}
The raw and pipeline reduced SALT observations are available from the SALT Data Archive ({\tt https://ssda.saao.ac.za}). The sky-subtracted, wavelength-calibrated and order-merged spectra will be made available from Data Central ({\tt https://datacentral.org.au}). 

\bibliographystyle{mnras}
\bibliography{ehe,cats}

\appendix
\renewcommand\thefigure{A.\arabic{figure}} 
\renewcommand\thetable{A.\arabic{table}} 
\section[]{SALT Classifications}
\label{s:app1}
{Table A1} 

shows the spectral classifications for {697}  stars in the SALT 
survey of helium-rich hot subdwarfs, including position (J2000), \textit{Gaia} G magnitude and  DR3 identifier {(columns 2 -- 4)},  up to one other familiar identifier {(column 5)}, spectral class {(column 6)}, {Drilling class (D13)} by number and errors {(columns 7 -- 12)}, and other catalogues containing the star {(column 13)}. 

\begin{table*}
\begin{center}
\caption{Spectral classifications for helium-rich hot subdwarfs and other stars observed with SALT.}
\label{a:classes}
\setlength{\tabcolsep}{2pt}
\begin{tabular}{r lcrl l  lrr lll l}
\hline
 \# & SALT & m$_{\rm G}$ & Gaia DR3 & Other & class & sp & lc & he & e$_{\rm sp}$  &  e$_{\rm lc}$ &  e$_{\rm he}$ &  catalogues \\
\hline
001 & J000331.6-164358 &  14.60 &  2414888694102153088 &  BPS CS 29517-0049 &  DB 13.0/8.1$^c$ &  -- & -- & -- & -- & -- & -- & PHL  EC  WD   \\
002 & J000346.7-483937 &  16.50 &  4977727221867267968 &  JL 152 &  sdO9VII:He39 &  0.90 & 7 & 39 &  0.07 & 1.5 &  1.3 &    \\
003 & J000431.0-242621 &  13.82 &  2336589275632410496 &  HE 0001-2443 &  sdB0.5VII:He40 &  1.05 & 7 & 40 &  0.01 & 0.0 &  0.4 &   SB  PHL  EC  BPS  GALEX   \\
004 & J001006.9-261256 &  12.78 &  2335246080444203520 &  TON S 144 &  sdO9.5VII:He37$^b$ &  0.95 & 7 & 37 &  0.01 & 1.4 &  0.3 &   SB  PHL  FB  EC  BPS  GALEX   \\
005 & J001235.6-422011 &  16.13 &  4995343592869448960 &   &  sdOC9.5VI:He25 Pb &  0.95 & 6 & 25 &  0.01 & 1.0 &  0.5 &   GALEX   \\
006 & J001853.2-315601 &  14.37 &  2316462680926136960 &  HE 0016-3212 &  sdBC0.2VI:He37$^b$ &  1.02 & 6 & 37 &  0.02 & 2.1 &  0.5 &   PHL  EC  GALEX   \\
007 & J002334.0+065647 &  13.91 &  2747736702353007488 &  PB 5919 &  CV ux &  -- & -- & -- & -- & -- & -- &  \\
\hline
\end{tabular}
\end{center}
\end{table*}
\addtocounter{table}{-1}
\begin{table*}
\begin{center}
\caption{contd.}
\setlength{\tabcolsep}{2pt}
\begin{tabular}{r lcrl l  lrr lll l}
\hline
 \# & SALT & m$_{\rm G}$ & Gaia DR3 & Other & class & sp & lc & he & e$_{\rm sp}$  &  e$_{\rm lc}$ &  e$_{\rm he}$ &  catalogues \\
\hline
066 & J041530.2-542158 &  14.85 &  4779256473879923072 &  HE 0414-5429 &  sdO8VII:He39$^b$ &  0.80 & 7 & 39 &  0.02 & 1.3 &  0.3 &   EC  GALEX   \\
067 & J042034.8+012040 &  12.29 &  3255780171819962496 &   &  sdOC9VII:He39$^b$ &  0.90 & 7 & 39 &  0.02 & 0.8 &  0.4 &    \\
068 & J042237.3-540849 &  14.03 &  4779307463731785088 &  LB 1721 &  sdOC9VII:He38$^b$ &  0.90 & 7 & 38 &  0.04 & 1.3 &  0.8 &   EC  GALEX   \\
069 & J042910.7-290248 &  14.08 &  4891033684953481600 &  EC 04271-2909 &  sdO8.5VI:He39$^b$ &  0.85 & 6 & 39 &  0.02 & 1.2 &  0.4 &   BPS  GALEX   \\
070 & J042910.8-041923 &  16.02 &  3204384497212064640 &   &  -sdO/DO- &  -- & -- & -- & -- & -- & -- &  \\
071 & J042932.9-473144 &  15.78 &  4789068103329323648 &  EC 04281-4738 &  sdOC6.5VII:He39$^b$ &  0.65 & 7 & 39 &  0.04 & 0.9 &  0.9 &   GALEX   \\
072 & J043213.8-164508 &  15.14 &  3171501952716992256 &  EC 04299-1651 &  DB 15.6/7.7 + dM$^d$ &  -- & -- & -- & -- & -- & -- & HE  GALEX  WD   \\
\hline
\end{tabular}
\end{center}
\end{table*}
\addtocounter{table}{-1}
\begin{table*}
\begin{center}
\caption{contd.}
\setlength{\tabcolsep}{2pt}
\begin{tabular}{r lcrl l  lrr lll l}
\hline
 \# & SALT & m$_{\rm G}$ & Gaia DR3 & Other & class & sp & lc & he & e$_{\rm sp}$  &  e$_{\rm lc}$ &  e$_{\rm he}$ &  catalogues \\
\hline
131 & J071436.6-323229 &  13.87 &  5603285806919711360 &   &  sdO2V:He09 &  0.20 & 5 &  9 &  0.05 & 0.8 &  0.3 &    \\
132 & J071549.7-540756 &  14.34 &  5491580506476906752 &   &  sdO3VII:He40$^b$ &  0.30 & 7 & 40 &  0.03 & 0.7 &  0.8 &    \\
133 & J071656.4-281944 &  13.20 &  5606816922877511424 &   &  sdB0VII:He21 &  1.00 & 7 & 21 &  0.01 & 0.5 &  0.5 &    \\
134 & J071741.2-465146 &  14.77 &  5510452287837529600 &   &  sdOC2V:He40 &  0.20 & 5 & 40 &  0.05 & 1.0 &  0.9 &    \\
135 & J072138.9+054422 &  14.86 &  3140865744837849216 &   &  sdO9VI:He34 &  0.90 & 6 & 34 &  0.02 & 0.5 &  0.5 &   GALEX   \\
136 & J072524.6-573041 &  14.48 &  5486262894585667456 &   &  sdBC0.5V:He38 &  1.05 & 5 & 38 &  0.02 & 1.1 &  0.5 &    \\
137 & J072951.1-281423 &  14.03 &  5611934840266445568 &   &  sdBN0VI:He29 &  1.00 & 6 & 29 &  0.02 & 1.3 &  0.6 &    \\
\hline
\end{tabular}
\end{center}
\end{table*}
\addtocounter{table}{-1}
\begin{table*}
\begin{center}
\caption{contd.}
\setlength{\tabcolsep}{2pt}
\begin{tabular}{r lcrl l  lrr lll l}
\hline
 \# & SALT & m$_{\rm G}$ & Gaia DR3 & Other & class & sp & lc & he & e$_{\rm sp}$  &  e$_{\rm lc}$ &  e$_{\rm he}$ &  catalogues \\
\hline
196 & J100818.5-594954 &  14.66 &  5258029809978171264 &   &  sdO5VI:He04 &  0.50 & 6 &  4 &  0.15 & 1.1 &  0.8 &    \\
197 & J101130.1-484059 &  13.70 &  5407246731987593984 &   &  sdO2VI:He12 &  0.20 & 6 & 12 &  0.04 & 0.5 &  0.5 &    \\
198 & J101421.1-565330 &  16.33 &  5258974457944841344 &   &  sdO4VIII:He08 &  0.40 & 8 &  8 &  0.01 & 1.3 &  0.7 &    \\
199 & J101827.0-401316 &  14.77 &  5417595953892026240 &   &  sdO5V:He03 &  0.50 & 5 &  3 &  0.02 & 0.5 &  0.3 &    \\
200 & J102646.0-323958 &  14.25 &  5448480651398698752 &   &  sdB0V:He11 &  1.00 & 5 & 11 &  0.02 & 2.2 &  0.6 &   GALEX   \\
201 & J102926.5-683321 &  13.78 &  5232542408868806144 &   &  sdOC5VI:He40 CO &  0.50 & 6 & 40 &  0.04 & 0.3 &  0.9 &    \\
202 & J103244.8-704312 &  14.02 &  5230573973813495808 &   &  sdO4VI:He02 &  0.40 & 6 &  2 &  0.02 & 0.5 &  0.4 &    \\
\hline
\end{tabular}
\end{center}
\end{table*}
\addtocounter{table}{-1}
\begin{table*}
\begin{center}
\caption{contd.}
\setlength{\tabcolsep}{2pt}
\begin{tabular}{r lcrl l  lrr lll l}
\hline
 \# & SALT & m$_{\rm G}$ & Gaia DR3 & Other & class & sp & lc & he & e$_{\rm sp}$  &  e$_{\rm lc}$ &  e$_{\rm he}$ &  catalogues \\
\hline
261 & J130547.7-473153 &  14.70 &  6086085004808015744 &   &  sdO6VII:He03 &  0.60 & 7 &  3 &  0.17 & 3.4 &  2.1 &    \\
262 & J130717.0-490046 &  14.25 &  6085009613716747776 &   &  sdOC6VI:He40 &  0.60 & 6 & 40 &  0.04 & 0.7 &  1.1 &    \\
263 & J130934.9-744250 &  13.41 &  5839006011691771648 &   &  sdO7.5VI:He39 &  0.75 & 6 & 39 &  0.05 & 0.5 &  1.0 &    \\
264 & J131315.1-720632 &  15.57 &  5840393324794400896 &   &  sdB0VII:He14 &  1.00 & 7 & 14 &  0.08 & 0.8 &  2.8 &    \\
265 & J131315.7-520452 &  13.72 &  6080503437111828864 &   &  sdO5IV:He05 &  0.50 & 4 &  5 &  0.03 & 0.6 &  0.2 &    \\
266 & J131607.4-701236 &  14.09 &  5843886507606067200 &   &  sdO8VI:He40 &  0.80 & 6 & 40 &  0.03 & 1.0 &  0.6 &    \\
267 & J131641.1-521956 &  15.69 &  6068452102478326912 &   &  sdB3V:He06 dib &  1.30 & 5 &  6 &  0.03 & 1.5 &  1.0 &    \\
\hline
\end{tabular}
\end{center}
\end{table*}
\addtocounter{table}{-1}
\begin{table*}
\begin{center}
\caption{contd.}
\setlength{\tabcolsep}{2pt}
\begin{tabular}{r lcrl l  lrr lll l}
\hline
 \# & SALT & m$_{\rm G}$ & Gaia DR3 & Other & class & sp & lc & he & e$_{\rm sp}$  &  e$_{\rm lc}$ &  e$_{\rm he}$ &  catalogues \\
\hline
326 & J145435.5-365618 &  16.45 &  6202023833031619456 &   &  sdO7.5VI:He01 &  0.75 & 6 &  1 &  0.29 & 2.0 &  1.4 &    \\
327 & J145538.5-384304 &  15.96 &  6198310438667511168 &   &  sdOC2VI:He40 &  0.20 & 6 & 40 &  0.06 & 1.4 &  1.2 &    \\
328 & J145757.4-070505 &  16.35 &  6333388770734074368 &  PG 1455-069 &  sdOC8.5VII:He40$^b$ &  0.85 & 7 & 40 &  0.09 & 0.1 &  1.6 &   GALEX   \\
329 & J145807.6-381422 &  14.85 &  6198338304416045824 &   &  sdOC3V:He39 &  0.30 & 5 & 39 &  0.04 & 0.4 &  0.9 &    \\
330 & J145858.9-413656 &  13.27 &  6100894052044889728 &   &  sdO4V:He04 &  0.40 & 5 &  4 &  0.02 & 0.6 &  0.1 &    \\
331 & J150043.1-323019 &  14.91 &  6205407820583007104 &   &  sdOC9VII:He39 &  0.90 & 7 & 39 &  0.02 & 1.1 &  0.8 &    \\
332 & J150206.4-045938 &  13.35 &  6335746093599431296 &  PG 1459-048 &  sdB1.5IV:He09 K+G &  1.15 & 4 &  9 &  0.02 & 0.0 &  0.9 &    \\
\hline
\end{tabular}
\end{center}
\end{table*}
\addtocounter{table}{-1}
\begin{table*}
\begin{center}
\caption{contd.}
\setlength{\tabcolsep}{2pt}
\begin{tabular}{r lcrl l  lrr lll l}
\hline
 \# & SALT & m$_{\rm G}$ & Gaia DR3 & Other & class & sp & lc & he & e$_{\rm sp}$  &  e$_{\rm lc}$ &  e$_{\rm he}$ &  catalogues \\
\hline
391 & J163554.6-591043 &  13.92 &  5831017372517593600 &   &  sdO6VI:He02 &  0.60 & 6 &  2 &  0.18 & 1.8 &  1.6 &    \\
392 & J163621.8-080636 &  11.45 &  4351051086327475456 &  LS IV -08 1 &  sdB3V:He13 dib &  1.30 & 5 & 13 &  0.03 & 2.3 &  0.9 &    \\
393 & J163828.3-245111 &  12.10 &  6047377209046344064 &   &  sdO9VI:He27 &  0.90 & 6 & 27 &  0.01 & 1.8 &  0.3 &    \\
394 & J163947.4-663151 &  13.28 &  5815599887346246144 &   &  sdON9.5VI:He36 &  0.95 & 6 & 36 &  0.01 & 1.4 &  0.2 &    \\
395 & J163958.4-604216 &  15.43 &  5830605541014576512 &   &  sdB0.5IV:He10 &  1.05 & 4 & 10 &  0.02 & 2.9 &  0.8 &    \\
396 & J164057.2-253434 &  15.57 &  6046526363151008128 &   &  sdB0.5V:He06 &  1.05 & 5 &  6 &  0.06 & 2.7 &  2.2 &    \\
397 & J164607.8-271718 &  15.38 &  6034028180126222592 &   &  sdB2VI:He14 &  1.20 & 6 & 14 &  0.02 & 1.4 &  1.0 &    \\
\hline
\end{tabular}
\end{center}
\end{table*}
\addtocounter{table}{-1}
\begin{table*}
\begin{center}
\caption{contd.}
\setlength{\tabcolsep}{2pt}
\begin{tabular}{r lcrl l  lrr lll l}
\hline
 \# & SALT & m$_{\rm G}$ & Gaia DR3 & Other & class & sp & lc & he & e$_{\rm sp}$  &  e$_{\rm lc}$ &  e$_{\rm he}$ &  catalogues \\
\hline
456 & J175411.9-265120 &  13.62 &  4063822880675094912 &   &  sdB0.2VII:He25 &  1.02 & 7 & 25 &  0.01 & 0.4 &  0.6 &    \\
457 & J175624.4-414608 &  15.43 &  5956800231430499584 &   &  sdB1.5V:He08 &  1.15 & 5 &  8 &  0.03 & 2.5 &  1.0 &    \\
458 & J175648.1-472246 &  13.89 &  5953840659677202816 &   &  sdO9.5VI:He27 &  0.95 & 6 & 27 &  0.03 & 0.9 &  1.2 &    \\
459 & J175731.7-683239 &  15.46 &  5811947687666193280 &   &  sdB0.2II:He05 &  1.02 & 2 &  5 &  0.02 & 1.8 &  0.7 &   GALEX   \\
460 & J175848.2-414950 &  15.04 &  5956763745729828864 &   &  sdO8.5VI:He38 &  0.85 & 6 & 38 &  0.02 & 1.7 &  0.8 &    \\
461 & J180106.5-694203 &  15.24 &  6432618874807952384 &   &  sdBC0V:He40 &  1.00 & 5 & 40 &  0.03 & 0.8 &  1.0 &   GALEX   \\
462 & J180218.6-131029 &  11.70 &  4150294311931498368 &  LS IV -13 2 &  sdO5VI:He02 dib &  0.50 & 6 &  2 &  0.04 & 0.6 &  0.2 &    \\
\hline
\end{tabular}
\end{center}
\end{table*}
\addtocounter{table}{-1}
\begin{table*}
\begin{center}
\caption{contd.}
\setlength{\tabcolsep}{2pt}
\begin{tabular}{r lcrl l  lrr lll l}
\hline
 \# & SALT & m$_{\rm G}$ & Gaia DR3 & Other & class & sp & lc & he & e$_{\rm sp}$  &  e$_{\rm lc}$ &  e$_{\rm he}$ &  catalogues \\
\hline
521 & J185417.3-455757 &  14.63 &  6710782255635813248 &   &  sdOC7VI:He40 &  0.70 & 6 & 40 &  0.03 & 1.0 &  0.7 &    \\
522 & J185505.3-625522 &  15.32 &  6439480858153283712 &   &  sdO3V:He10 &  0.30 & 5 & 10 &  0.10 & 2.4 &  3.9 &    \\
523 & J185505.7-294131 &  15.25 &  6761614263052184064 &   &  sdO9VI:He13 &  0.90 & 6 & 13 &  0.03 & 0.6 &  1.3 &    \\
524 & J185515.6-621213 &  14.33 &  6439577198562118656 &   &  sdO9VII:He39 &  0.90 & 7 & 39 &  0.01 & 1.2 &  0.5 &    \\
525 & J185522.2-090725 &  11.11 &  4203620836396895104 &  LS IV -09 34 &  sdB2.5III:He18 &  1.25 & 3 & 18 &  0.14 & 0.7 &  0.6 &    \\
526 & J185611.6-090748 &  14.87 &  4203578883153216256 &   &  sdBN0VI:He34 &  1.00 & 6 & 34 &  0.02 & 0.9 &  0.7 &    \\
527 & J185630.2-183130 &  15.39 &  4086573764816125824 &   &  sdO9.5VI:He21 &  0.95 & 6 & 21 &  0.05 & 1.6 &  2.4 &    \\
\hline
\end{tabular}
\end{center}
\end{table*}
\addtocounter{table}{-1}
\begin{table*}
\begin{center}
\caption{contd.}
\setlength{\tabcolsep}{2pt}
\begin{tabular}{r lcrl l  lrr lll l}
\hline
 \# & SALT & m$_{\rm G}$ & Gaia DR3 & Other & class & sp & lc & he & e$_{\rm sp}$  &  e$_{\rm lc}$ &  e$_{\rm he}$ &  catalogues \\
\hline
586 & J200439.2-424627 &  15.55 &  6686054312141146112 &   &  sdO3VII:He03 rvv &  0.30 & 7 &  3 &  0.03 & 0.5 &  0.2 &   GALEX   \\
587 & J200624.2-132759 &  14.29 &  6878854122751996288 &   &  sdO9.5VI:He33 &  0.95 & 6 & 33 &  0.02 & 0.3 &  0.9 &    \\
588 & J200741.2-425239 &  16.00 &  6686182577044658176 &   &  sdO9.5VII:He28 &  0.95 & 7 & 28 &  0.02 & 2.3 &  0.9 &   GALEX   \\
589 & J200813.1-132649 &  14.19 &  6878863880917504384 &   &  sdB3II:He10 &  1.30 & 2 & 10 &  0.03 & 1.5 &  0.9 &   GALEX   \\
590 & J200849.1-510411 &  15.82 &  6666280523228538112 &  MCT 2005-5112 &  sdO2V:He12 &  0.20 & 5 & 12 &  0.11 & 0.7 &  3.3 &    \\
591 & J200854.3-223544 &  15.22 &  6853439186350811008 &   &  sdB1.5V:He08 &  1.15 & 5 &  8 &  0.03 & 0.0 &  1.1 &   GALEX   \\
592 & J201118.7-315606 &  16.18 &  6700716918173536768 &  EC 20081-3205 &  sdON5VII:He40 &  0.50 & 7 & 40 &  0.01 & 2.9 &  0.4 &   GALEX   \\
\hline
\end{tabular}
\end{center}
\end{table*}
\addtocounter{table}{-1}
\begin{table*}
\begin{center}
\caption{contd.}
\setlength{\tabcolsep}{2pt}
\begin{tabular}{r lcrl l  lrr lll l}
\hline
 \# & SALT & m$_{\rm G}$ & Gaia DR3 & Other & class & sp & lc & he & e$_{\rm sp}$  &  e$_{\rm lc}$ &  e$_{\rm he}$ &  catalogues \\
\hline
651 & J220839.9-520323 &  14.35 &  6509903272241240960 &   &  sdO9VI:He13 &  0.90 & 6 & 13 &  0.01 & 1.7 &  0.7 &    \\
652 & J221032.3-454056 &  13.43 &  6566766440217253120 &   &  sdB0.2VI:He09 &  1.02 & 6 &  9 &  0.01 & 0.3 &  0.4 &    \\
653 & J221457.5-634145 &  14.51 &  6405416373015926656 &  BPS CS 22956-0090 &  sdO9VII:He40$^b$ &  0.90 & 7 & 40 &  0.02 & 1.2 &  0.4 &   EC  GALEX   \\
654 & J221604.5-171947 &  14.61 &  6825916554879185664 &  BPS CS 22892-0051 &  sdOC7VII:He40$^b$ &  0.70 & 7 & 40 &  0.03 & 1.2 &  0.6 &   EC  GALEX   \\
655 & J221628.4-002113 &  14.08 &  2678666962443647232 &  PG 2213-006 &  sdO7.5VI:He39 &  0.75 & 6 & 39 &  0.02 & 0.9 &  0.5 &   PB  FBS  GALEX  SDSS   \\
656 & J221656.0-643150 &  13.30 &  6404530097924835968 &  BPS CS 22956-0094 &  sdBC0.5V:He24 &  1.05 & 5 & 24 &  0.02 & 1.9 &  0.7 &   EC   \\
657 & J221720.1-384730 &  16.41 &  6573749640427200384 &   &  sdO2VII:He14 &  0.20 & 7 & 14 &  0.09 & 3.0 &  3.8 &   GALEX   \\
\hline
\end{tabular}
\end{center}
\end{table*}

\begin{table*}
    \centering
    \begin{tabular}{P{175mm}}
    \parbox{175mm}{Fundamental data: $\alpha_{2000}$, $\delta_{2000}$, $m_G$: \citet{gaia23.dr3}.} \\[1mm]
    \parbox{175mm}{Associated nebulae:
    1: NGC\,246 \citep{Cat.NGC}, 
    2: JeWeKi 1 \citep{jeffery23a}.
    } \\
\end{tabular}
\end{table*}

\label{lastpage}
\end{document}